\documentclass[a4paper, 12pt]{ article}                   
\usepackage{graphicx}

\begin{document}

\title{Subcritical Star Formation in Galactic Discs:  \\ the role of Dark Matter and Magnetic Field}

\author{ Khalil Chamcham \\
\thanks{E-mail: khalil.chamcham@physics.ox.ac.uk}
 University of Oxford, Astrophysics Sub-Department, \\ Beecroft Institute for Particle Astrophysics \& Cosmology \\ 
 1, Keble Road, OX1 3RH, Oxford, UK}

%
%

\maketitle

\label{firstpage}

\begin{abstract}
We discuss the occurrence of subcritical star formation in the outskirts of some galactic discs and across LSB galaxies, contrary to the picture that star formation happens only when the gas surface density is above a critical threshold density. This raises the question of whether the Toomre Q-criterion for gas alone is valid for correlating the gas distribution to star formation or rather some effective Q-parameter, taking account of components such as stars, dark matter and magnetic fields, is more representative of the correlation between gas instability and star formation activity. 
As a potential candidate, we investigate the role of dark matter in triggering subcritical star formation, particularly at the outskirts where dark matter is dominant. Indeed, our axisymmetric analysis favours the picture that dark matter contributes to disc instability at the subcritical regime where gas seems stable when dark matter is neglected. 

\end{abstract}

\section*{keywords}
Galaxy: disc; galaxies: dark matter, magnetic fields; ISM: magnetic fields; star formation: subcritical. 

\newpage

\section{Introduction}
\label{sec intro}

Study of the outskirts of galaxies gained strong interest because of the peculiar behaviour of star formation in these regions (Hunter et al. 2016, Elmegreen \& Hunter 2016) as compared to the standard picture formed from the study of the inner regions of galaxies (Kennicutt 1989, Kennicutt \& Evans 2012).   

\par
There have been observational arguments in favour of star formation (SF) being controlled by the 2D Toomre $Q$-parameter (\ref{toomreq}) in its standard form for a galaxy gas component (Kennicutt 1989, Martin \& Kennicutt 2001): i.e. SF switch on when $Q_g < 1$ and off when $Q_g > 1$. 
The standard Toomre criterion (Toomre 1964) for the stability of a thin gaseous disc considered as a continuous fluid is expressed as

\begin{equation}
Q_{g} = \frac{\kappa c_g}{\pi G \Sigma_g} = \frac{\Sigma_c}{\Sigma_g} \;\; ,
\label{toomreq}
\end{equation}

\noindent
with $\kappa$ the epicylic frequency, $c_g$ the gas sound speed, and $G$ the gravitational constant. The gas critical density is obtained for $Q_g = 1$; discs are unstable for $Q_g < 1$ and stable for $Q_g > 1$. 

In this work we discuss whether this criterion or a modified version of it, including dark matter (DM) and magnetic field (MF), should be considered in order to account for a realistic dynamical picture. This is motivated by the examples of some irregular and LSB galaxies which are subcritical across their entire disc (i.e. SF is on with $Q_g > 1$) and at the same time  are known for being up to 10 times more dominated by DM than spiral galaxies (Hunter et. al. 1998, Kennicutt 1989, Martin \& Kennicutt 2001). This might also be the case in the outer regions of discs, far from the optical radius, where DM dominates (Ferguson et al. 1998, Leli\`evre \& Roy 2000b).

\par
Star formation remains a key process to understanding galaxy formation and evolution, as well as the matter distribution at large scales. Identifying the mechanisms controlling the process of SF provides insights into the dynamics of galaxies and set constraints on the initial conditions of their formation (Slyz et al. 2005). Authors investigating the connection between SF and the properties of the ISM have drawn an empirical link between gravitational radial instability of disc galaxies and SF activity (Kennicutt 1989, Martin \& Kennicutt 2001, Wang \& Silk 1994, Elmegreen \& Hunter 2006, Schaye 2004), where the Toomre $Q$-criterion or equivalently the critical gas surface density, $\Sigma_c$, is identified as a parameter regulating the SF activity in H\,{\sc ii} regions. This threshold for SF requires that the gas surface density, $\Sigma_g$, must be larger than the critical gas surface density for SF to occur (Zasov \& Simakov 1988, Chamcham, Pitts \& Tayler 1993). 

\par
However, the slow evolution of SF in the Milky Way may suggest that SF evolves along the line of marginal stability where $\Sigma_g \sim \Sigma_c$ ($Q_g \sim 1$). Kennicutt argues that marginal stability is a universal trend in disk galaxies and that the existence of a SF threshold also implies that there is a galactic radius, R\,{\sc ii}, outside which SF is cut off. But this seems to be questioned by observations that show the existence of knots of stars far out at the outskirt of dwarf irregular (dIrr) galaxies where gas is predicted to be stable (Hunter et al. 2016).  
This leaves us with few questions: 1) what are the physical processes regulating SF?; 2) how do these physical processes combine to drive SF along the line of marginal stability? Otherwise speaking, is SF fine-tuned?; 3) is the threshold density, $\Sigma_c$, a realistic parameter or is there an effective critical density that encompasses more of the dynamics of SF?  
%

\par
Hunter et al. (2016) analysed FUV images of the LITTLE THINGS of the nearby dwarf irregulars (dIrr) and Blue Compact Dwarf galaxies and found that the gas in these regions is stable against collapse into star-forming clouds if scrutinised by the standard 2D Toomre criterion (\ref{toomreq}): the presence of FUV emission in the outer disk suggests a continuing SF with radius (eg. young star clusters and OB associations). In addition, most regions are found around an H\,{\sc i} surface density of $1 \; {\rm M_{\odot} pc^{-2}}$ suggesting the existence of a threshold-like for SF. 

\par
More observations showed that the Toomre parameter $Q_{g}$ may not be a strong indicator of SF in some normal disc galaxies (Thornley \& Wilson 1995, Ferguson et al. 1998, Ferguson et al. 1999), and even in irregulars (Hunter, Elmegreen \& Baker 1998) and Low Surface Brightness galaxies (van der Hulst et al. 1993). H$\alpha$ mapping (Ferguson et al. 1998, Leli\`evre \& Roy 2000a) has shown evidence of star-forming regions only a few mega-years old beyond two optical radii in some galaxies in regions where  SF can be subcritical. This has been corroborated by GALEX observations of the outer disc of M83 in the UV (Thilker et al. 2005). 
There are even cases where the gas surface density lies well below the critical density throughout most of the disc, i.e. the cases of M33 (Kennicutt 1989), NGC 2403 (Thornley \& Wilson 1995) and Sextan A (Hunter \& Plummer 1996). 

\par
The complexity of the physics of the ISM in its relation to the SF process indicates that the Toomre criterion may not encompass all the dynamics that drives star formation, i.e. that gas turbulence, feedback and self-gravity are not the only actors at play. For instance, the fact that the lifetime of molecular clouds is substantially larger than their free-fall timescale suggests that gravitational instabilities are only a necessary phase for cloud formation, but the rate at which the existing clouds will turn into stars depends on conditions like the ability of clouds to survive cloud-cloud collisions, disruption by shock waves from SNe (Silk 2000, Slyz et al. 2005) and stellar winds (Norman \& Silk 1980), the complexity of the microphysics and the role of MF (Elmegreen 1991, Mestel 1999) and DM.

\par
The flat rotation curve of disc galaxies suggests that dark matter from the halo constitutes a major component of disc galaxies (Bosma 1999 (a)-(b)); in particular, it dominates the outer regions of galaxies and this suggests that gas dynamics in these regions can differ appreciably  from the inner regions where baryons dominate. Also the effect of DM is particularly important in irregular galaxies where its contribution can be up to 10 times higher than in spiral galaxies (Hunter et al. 1998, Carignan \& Freeman 1985, Kennicutt 1989). 

\par
This paper is organised as follows: in sections \ref{subsec Gunperturb2} and \ref{subsec Dispers} we discuss the stability conditions of a self-gravitating gas layer in presence of DM and MF. In section \ref{sec GS stability} we discuss the stability criterion for a realistic disc with stars, gas, DM and MF. In section \ref{sec model} we discuss our model parameters. Finally in Section \ref{sec effectiveparam} we discuss the effective 'Kennicutt' $\alpha$-parameter, the effective Q-criterion for discs and its associated critical surface density. 
 
\section{ The stability criterion }
\label{sec Gstability}

We determine the stability criterion of a multi-component galactic disc made of gas, stars, MF and including the DM contribution from the dark halo by studying its response to an external perturbation (Jog \& Solomon 1984, BT08, Elmegreen 1987, 1991, 1995). The disc is considered an axisymmetric, self-gravitating thin layer. The assumption of axisymmetry is to be used with caution: for instance MF in an axisymmetric disc resist instability while they favour instabilities in spirals particularly at low shear (Elmegreen 1994). However this should be a good approximation for representing the averaged stability properties of the disc. 

\par
The baryonic disc component, of scale height $z_d$, is assumed to lie at the equatorial plane of the spherical halo DM made of WIMPs. We consider the amount of halo DM embedded within $2 z_d$ and we focus on the response of the whole system (baryons + DM + MF) to a radial perturbation within the disc. The aim is to show that gravitational instability driven by luminous baryons alone is not enough to drive SF: in his seminal work Kennicutt (1989) showed that SF is happening only in regions where the gas surface density is above some critical value ($\Sigma_g \geq \Sigma_c$) or equivalently the Toomre criterion $Q_g \leq 1$, but this result could not be generalised to all galaxies. On the other hand, although MHD instabilities are an efficient mechanism to drive turbulence (Sellwood and Balbus, 1999) they may not be enough to drive the instabilities that trigger SF since magnetic fields, in some circumstances, stabilise the disc by thickening it (eg. magnetic buoyancy), hence shutting down SF. 

\par
We argue that the combination of Alfv\`enic turbulence and a hidden DM component that add to the self-gravity of the luminous baryonic component can contribute to drive SF: i.e. this effect manifests itself prominently in regions of the disc where SF is observed to be subcritical ($\Sigma_g < \Sigma_c$), hence deviating from the standard picture established by Kennicutt's data.   

\par
In the following sections, we first develop the equilibrium and the linearized equations of a differentially rotating sheet of gas in presence of DM halo and under the influence of MF (see details in Appendices A \& B), then we generalise our discussion to include the contribution of stars.

\subsection{Gas disc with MF and DM}
\label{subsec Gunperturb2}

We first treat the simple case of a thin circular gas disc (see Appendix A for details).
Without loss of generality we consider that the gas is barotropic and thus its sound speed is defined as $c_g^{2} = {dP}/{d\Sigma} $.


\par
The radial and azimuthal components of the Euler equation lead successively to 

\begin{eqnarray}
\frac{v_{0\varphi}^2}{R} & = & \frac{1}{\Sigma_{g}}\frac{d P_0}{d R} 
+ \frac{1}{\Sigma_{g}}\frac{d}{d R} (\frac{B_{0\varphi}^2}{8\pi})
+ \frac{1}{4 \pi \Sigma_{g}}\frac{B_{0\varphi}^2}{R}
+ \frac{d \Phi_0}{d R}     \nonumber  \\                       
                                     & = & ( c_{g}^2 + \frac{1}{2} c_{A}^2  ) \frac{d}{dR} \ln \Sigma_{g} 
+ \frac{ c_{A}^2 }{R} + \frac{d \Phi_0}{d R}  
\label{reul0}
\end{eqnarray}

\noindent
and

\begin{equation}
\frac{B_{0R}}{R} \frac{\partial}{\partial R} \left(R B_{0\varphi} \right) = 0 
\;\; .
\label{feul0}
\end{equation}

\noindent
The radial component of the field equation is trivial and its azimuthal component is 

\begin{equation}
B_{0R} \left( \frac{\partial v_{0\varphi}}{\partial R} - \frac{v_{0\varphi}}{R}
\right) = 0  \;\; .
\label{fmag0}
\end{equation}

\noindent
Finally, the divergence of the field satisfies

\begin{equation}
 \frac{\partial ( R B_{0R} )}{\partial R} = 0  \;\; ,
\label{div0}
\end{equation}

\noindent
the continuity equation is trivial, and the Poisson equation can be written

\begin{equation}
\frac{1}{R} \frac{\partial}{\partial R}\left( R \frac{\partial \Phi_0}
{\partial R} \right) + \frac{ \partial^2 \Phi_0 }{\partial z^2} =
4 \pi G \rho_{tot} \;\; ,
\label{Pois2}
\end{equation}

\noindent
hence the vertical gravitational acceleration within a latitude $z_i$

\begin{equation}
 | K_z (R,z_i) | = \frac{ \partial \Phi_0 }{\partial z} |_{z_i} = 2 \pi G  \Sigma (R,z_i) - \int_0^{z_i} dz' \frac{1}{R} \frac{\partial}{\partial R}\left( R \frac{\partial \Phi_0}
{\partial R} \right)
\label{verticalforce}
\end{equation}

\noindent
which is determined by analysis of stellar kinematics.

\par
Most of the authors ignore the integral in equation (\ref{verticalforce}), assuming that the rotation curve is flat at all latitudes (see Read 2014 for a review). This approximation is valid as long as $z_i \sim 1 \; {\rm kpc}$ (Kuijken \& Gilmore 1989). However McFee et al. (2015) argue that this is a bad approximation as the rotation velocity is not flat above the plane of the disc. Bovy \& Tremaine (2012) argue that at higher latitudes this approximation can lead up to 20\% uncertainty for $z_i \sim 4 \; {\rm kpc}$. 
This issue adds up to the uncertainty of the measurement of the total surface density of the disc - and particularly the DM contribution - which remains a much debated topic (Bahcall et al. 1992, Kuijken \& Gilmore 1989, Read 2014). On the other hand it is worth paying attention to this approximation when one considers the effect of the thickness on the disc dynamics.
%

\par
Because the disc is differentially rotating, equation (\ref{fmag0}) implies that the radial component of the MF $B_{0R} = 0$. The fact that only the azimuthal component, $B_{0\varphi}$, of the MF survives is in accordance with observations which show that in the Milky Way, the mean direction of $ \vec{B}_{0} $ lies within the disc (Parker 1966, Ruzmaikin et al. 1988, figure 3.19 of Binney \& Merrifield 1998). Also studies of M51, M31 and M33 show that  the MF follows the spiral arms (Beck et al. 1996, Ruzmaikin et al. 1990, Moss et al. 1998, papers in Berkhuijsen et al. 2000, Neininger \& Horellou 1996).

\par
The second line of equation (\ref{reul0}) derives from the assumption that $B_{0\varphi}^{2} (R) \propto \Sigma_g (R)$ (i.e. $b_{0\varphi}^{2} (R) \propto \rho_g (R)$ - see Appendix A). This implies that the Alfv{\`en speed $c_{A}$ is constant throughout the disc: eg. Sellwood \& Balbus (1999) argue that the constancy of the Alfv\`en speed is a consequence of flux freezing . This can be modeled as

\[
b_{0\varphi}^{2} (R) = \frac{b_0^2}{\rho_0} \rho_g (R) \;\; {\rm or} \;\;    
B_{0\varphi}^{2} (R) = \frac{b_0^2}{\rho_0} \Sigma_g (R)  \; . 
\]

\noindent
Hence

\[
c_{A}^{2} =  \frac{ b_{0}^2 }{4 \pi \rho_0} \; , \; {\rm with} \;
\rho_0 = \rho_g (R_0) \; {\rm and} \;
b_0 = b_{0\varphi} (R_0) \; .
\]

\par
At the solar neighbourhood, $R_0 = 8.5 \; {\rm kpc}$, we use the volume density of the ISM $\rho_0 = 0.05 \; {\rm M_{\odot} pc^{-3} } \simeq 3.38 \; 10^{-24} \; {\rm g cm^{-3}}$ for an associated surface density $\Sigma_g (R_0) = 13 \; {\rm M_{\odot} pc^{-2}}$ (BT08); McKee et al. (2015) use  
$\rho_0 = 0.041 \pm 10 \% \; {\rm M_{\odot} pc^{-3} } \simeq 2.77 \; 10^{-24} \pm 10 \% \; {\rm g cm^{-3}}$ for the total mid-plane density of the ISM. For a magnetic field $b_0 \simeq 5 \; {\rm \mu Gauss}$ the Alfv\`en speed $c_A \sim 8 \; {\rm km s^{-1}}$ (and $c_A \simeq 6 \; {\rm km s^{-1}}$ for $b_0 = 4 \; {\rm \mu Gauss}$), a value comparable to the gas sound speed $c_g \sim 6 \;{\rm km s^{-1}}$ generally adopted.

\par
The gas sound speed and the Alfv\`en speed being small compared to the rotation speed $V_c \sim 200 \; {\rm km s^{-1}}$, equation (\ref{reul0}) can be reduced to

\begin{equation} 
v_{0\varphi}^{2} = V_c^{2} = R \left( \frac{\partial \Phi_{0}}{\partial R} \right)_{z=0} \;\; .
\label{rotg}
\end{equation}
%

\noindent
In presence of gas alone, the scale height of the disc is

\[z_{g0} = \frac{ c_g^2 }{\pi G \Sigma_{g} }  \;\;\; , \]

\noindent
while in the presence of MF and DM the scale height is

\begin{equation}
z_g =  \frac{ 1 + \alpha_m^2 }{ 2 \Gamma_g } z_{g0}  \; ,
\label{gthick}
\end{equation}

\noindent
with  \[   \alpha_m =  \sqrt{1 + \frac{c_A^2}{c_g^2}  }  \;\; , \;\;
              \Gamma_g = 1 + \frac{\Sigma_{dmg}}{\Sigma_g}  \; ,  \]

\noindent
and $\Sigma_{dmg} = \rho_{h} (R, z=0) 2 z_g $ the amount of halo DM embedded within the gas layer of thickness $2 z_g$ for a given density profile $\rho_h$ of the spherical halo.
This shows that MF oppose the gravitational collapse of the gas by `thickening' the disc while DM tends to reduce the disc thickness hence privileging disc instability. The neat result will depend on the relative strength between magnetic buoyancy and the gravitational pull of DM.  
%
%
%
%
%
%

\begin{figure}
\includegraphics[height=100mm, width=170mm,angle=0]{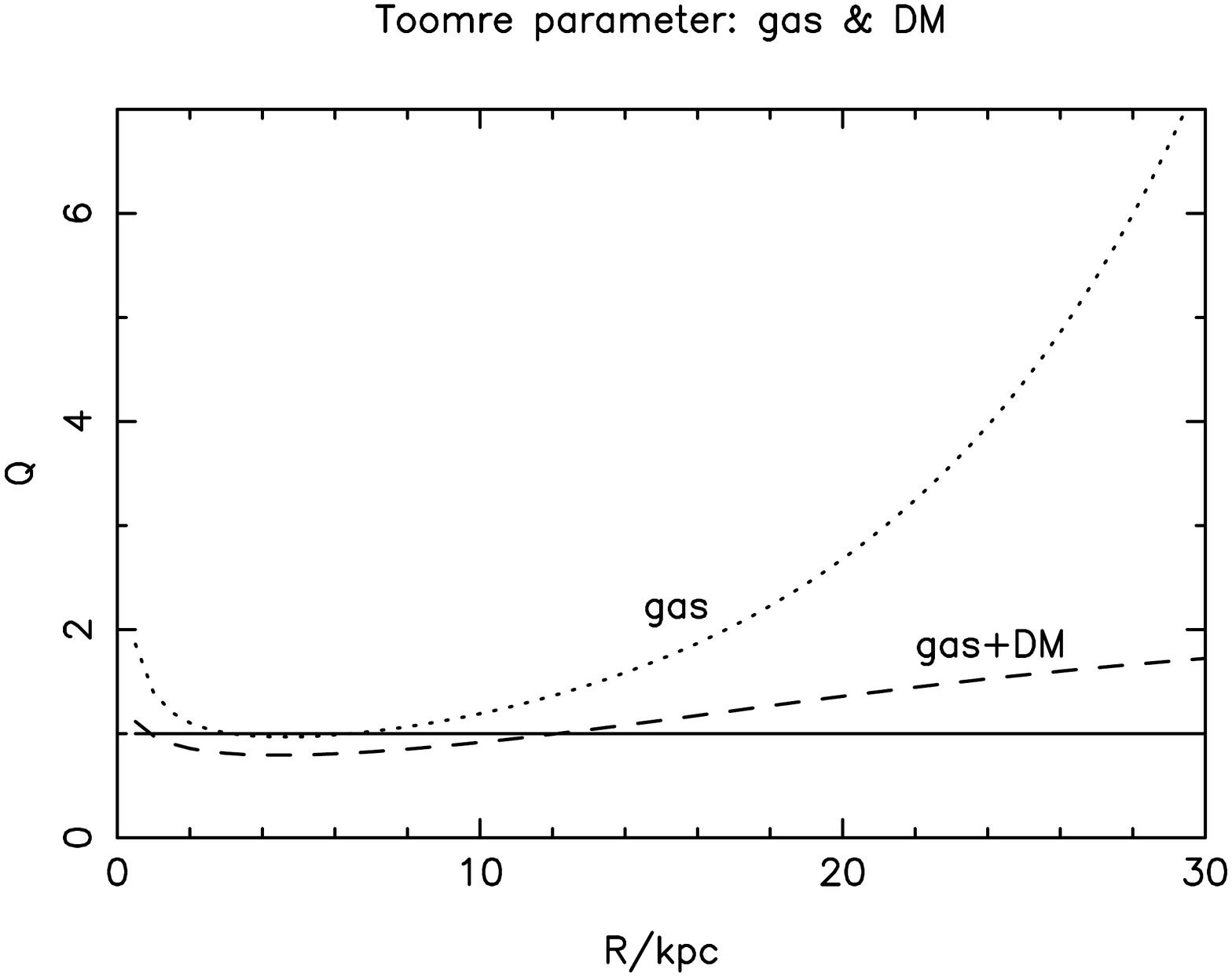}
\caption{ The standard Toomre parameter $Q_g$ (dot) and modified $Q_{ge}$ (dash) with the DM effect for gas only disc. The horizontal line shows the line of marginal stability.}
\label{Q-gasonly}
%
\includegraphics[height=100mm, width=170mm,angle=0]{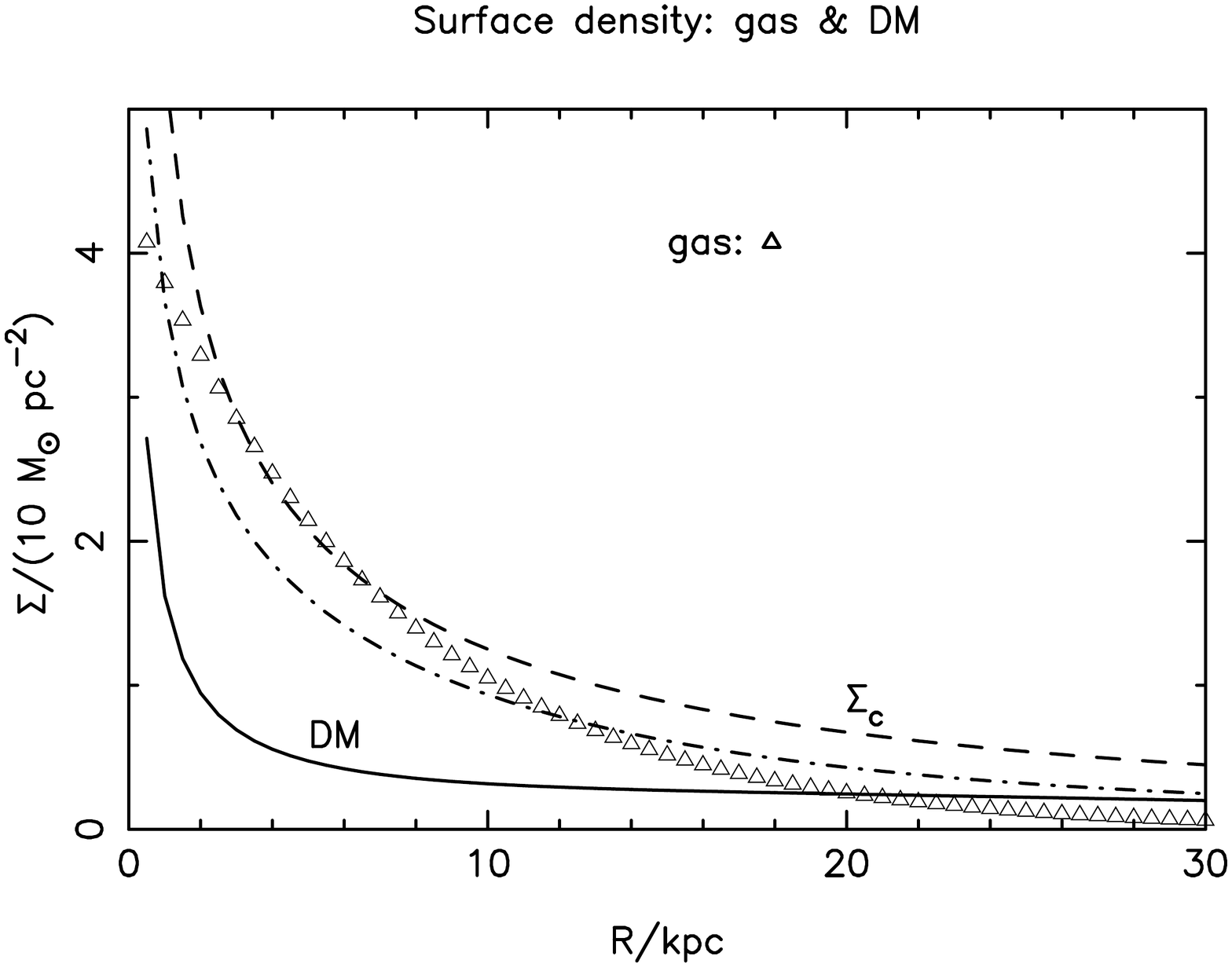}
\caption{Surface density $\Sigma_c$ (dash), $\Sigma_{ce}$ (\ref{gcriteff}) (dash-dot), gas (triangle) and DM (continuous).}
\label{Sdens-gasonly}
\end{figure}

\subsection{The modified Toomre criterion with MF and DM}
\label{subsec Dispers}

\indent
From the set of the linearised equations developed in Appendix B, we deduce the following homogeneous equations to solve for the amplitude of the perturbations $q_1$ in (\ref{sol})

\begin{equation} 
- i \omega v_{R1} - 2 \Omega v_{\varphi 1} = 
- ik \frac{c_{g}^{2}}{\Sigma_{g}} \Sigma_{g1} - ik \Phi_{1}  
- ik \frac{B_{0\varphi}}{4 \pi \Sigma_{g}} B_{\varphi 1}  \;\; ,   
\label{eq1}
\end{equation}

\begin{equation}
- i \omega v_{\varphi 1} - 2 B v_{R1} = \frac{1}{4 \pi \Sigma_{g}}
\frac{1}{R}\frac{\partial}{\partial R}(R B_{0\varphi}) B_{R1}   \;\; .
\label{eq2}
\end{equation}

\par
The radial component of the field equations shows that the component $B_{R1} = 0$,
hence reducing the azimuthal component to

\begin{equation}
k B_{0\varphi} v_{R1} - \omega B_{\varphi 1} = 0  \;\; .
\label{eq3}
\end{equation}

\noindent
The continuity equation gives

\begin{equation}
k \Sigma_{g} v_{R1} -  \omega \Sigma_{g1} = 0  \;\; .
\label{eq4}
\end{equation}

\noindent
Finally, solution of the above set of equations leads to the following 
dispersion relation 

\begin{equation}
\omega^{2} = \kappa^{2} - 2 \pi G \Gamma_g \Sigma_{g} k + c_{t}^{2} k^{2}  \;\;\; ,
\label{disprg}
\end{equation}

\noindent
which shows that the role of MF is to increase turbulence, hence increasing the gas sound speed $c_g$ to an effective value $c_t = \alpha_m c_g$. On the other hand DM makes the self-gravitating disc of surface density
$\Sigma_g$ behave like a disc of effective surface density $ \Sigma_{ge} = \Gamma_g \Sigma_g$. This can be interpreted as DM coupling to baryonic matter in a way that each gas particle of mass $m_g$ behaves like a particle with an effective mass $\Gamma_g m_g$.  

\par
\noindent
The epicyclic frequency $\kappa(R)$ is defined as (BT08)

\begin{equation}
\kappa^{2} (R) =  -4 B \Omega = \frac{1}{R^3} \frac{d}{dR}(R^2 V_{c}^2)  \;\; ,
\label{epicycl1}
\end{equation}

\noindent
where $\Omega(R) = V_{c} (R) / R $ is the circular velocity and $B$ the Oort `constant' 

\begin{equation}
 B(R) = - \frac{1}{2} \left\{ \frac{d(R\Omega)}{dR} + \Omega \right\} =
   - \frac{1}{2} \frac{1}{R} \frac{ d(R V_{c}) }{dR} \;\; .
\label{oort}
\end{equation}

\noindent
The second Oort `constant' $A$ is defined as

\begin{equation}
A =  - \frac{1}{2} R \frac{d}{dR} \left( \frac{V_{c}}{R} \right) \; ,
\end{equation}

\noindent
such that $\kappa(R_0) = \sqrt{-4B(A - B)}$.
Typical values of the Oort constants at $R_0$ are $ A = 14.5 \; \pm \; 1.5 \; {\rm km \; s^{-1} \; kpc^{-1} }$, 
$ B = -12.5 \; \pm \; 2 \; {\rm km \; s^{-1} \; kpc^{-1} } $, with 
$ A - B = 27 \; \pm \; 1.5 \; {\rm km \; s^{-1} \; kpc^{-1} }  = V_c/R$ (Dehnen \& Binney, 1998). However, these values will not fit a gas disc as they account for the total baryonic mass (gas + star) not gas only - i.e. the epicyclic frequency for a gas disc is lower than that of a realistic disc galaxy.
%

\begin{figure}
\includegraphics[height=100mm, width=170mm,angle=0]{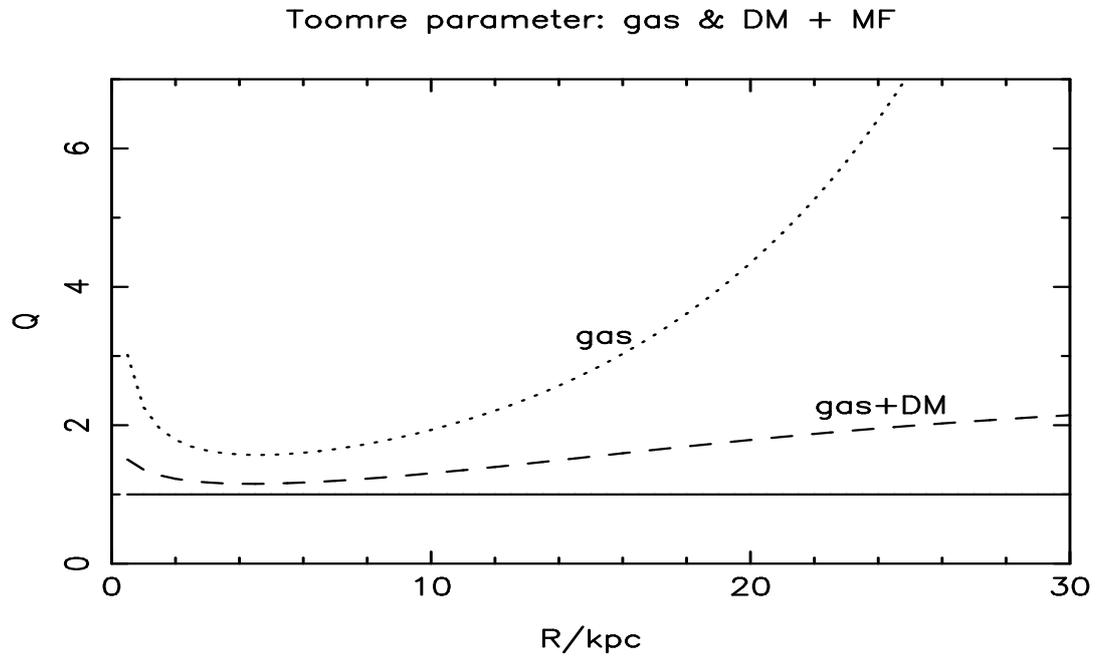}
\caption{ Identical to figure \ref{Q-gasonly} but with the effect of MF.
}
\label{Q-gasonlyMF}
\end{figure}

\begin{figure}
\includegraphics[height=100mm, width=170mm,angle=0]{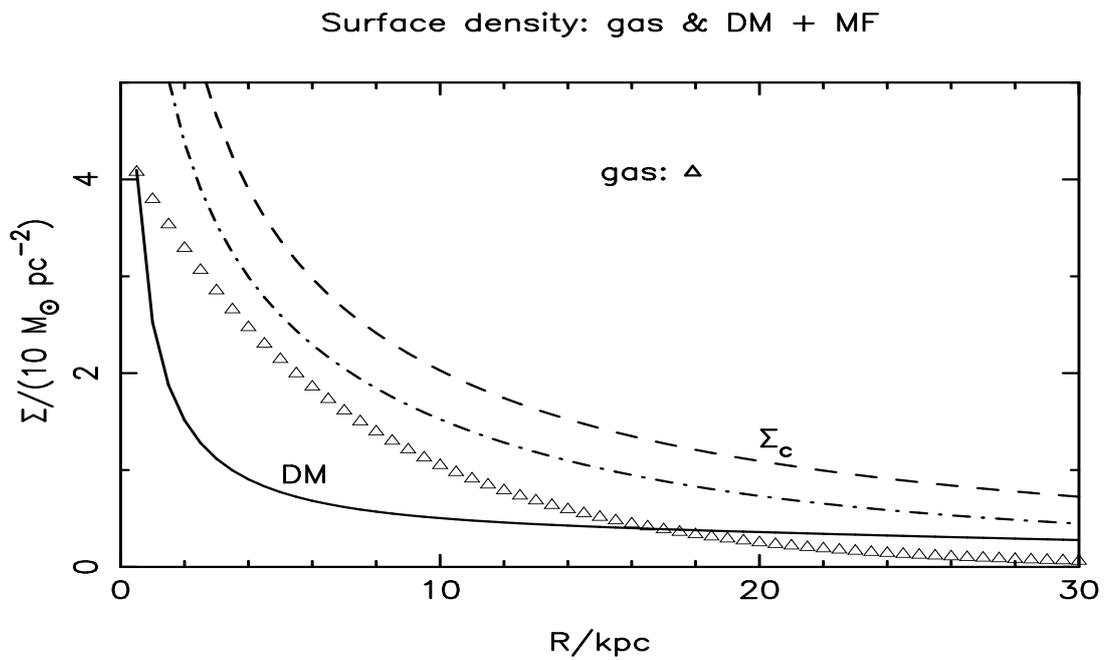}
\caption{ Identical to figure \ref{Sdens-gasonlyMF} but with the effect of MF.
}
\label{Sdens-gasonlyMF}
\end{figure}

\par
For the system to be unstable, we need the instability to grow with time, therefore $\omega^{2}$ should be a real negative number. In this case equation (\ref{disprg}) has a solution only if the quantity $Q_{ge}$ satisfies the inequality

\begin{equation}
Q_{ge} = \frac{ \kappa c_t }{ \pi G \Gamma_g \Sigma_{g} } = \frac{ \alpha_m }{\Gamma_g} Q_g \leq 1  \;\; ,
\label{toomre-gmfdm}
\end{equation}

\noindent
which is the modified Toomre criterion for the instability of a differentially rotating thin gaseous layer under the influence of DM and a MF parallel to the plane of the layer - $Q_g$ is the standard Toomre criterion defined in equation (\ref{toomreq}). Star formation will be regulated by a negotiation between the stabilising strength of MF (MHD turbulence) combined with gas turbulence and the amount of DM present in the disc that adds up to the self-gravity of baryonic matter. The effective critical surface density associated to criterion (\ref{toomre-gmfdm}) is

\begin{equation}
\Sigma_{ce} = \alpha_{m} \Sigma_c - \Sigma_{dmg} \; .
\label{gcriteff}
\end{equation}

\par
Using the model described in section \ref{sec model}, the DM surface density decreases sharply until the solar position where it settles at $\Sigma_{dmg} (R_0) \sim 3.4 \; {\rm M_{\odot} pc^{-2}}$ at the solar neighbourhood and then decreases very slowly up to about $ 2 \; {\rm M_{\odot} pc^{-2}}$ at 30 kpc (fig. \ref{Sdens-gasonly}). The gas component dominates DM in the inner region up to about 20 kpc where DM becomes relatively dominant outwards. However, as MF thickens the disc, the DM contribution starts dominating the gas component outwards much earlier at about 17 kpc; it is much higher at the solar neighbourhood, $\Sigma_{dmg} (R_0) \sim 5.5 \; {\rm M_{\odot} pc^{-2}}$, then decreases steadily to about $ 2.8 \; {\rm M_{\odot} pc^{-2}}$ at 30 kpc (fig. \ref{Sdens-gasonlyMF})

\par
Figures (\ref{Q-gasonly} \& \ref{Sdens-gasonly}) show that, when using the standard Toomre criterion and the associated threshold density, the gas disc is marginally stable (i.e. $Q_g \sim 1$ and $\Sigma_g \sim \Sigma_c$) within a small inner area between $3 - 7 \; {\rm kpc}$ and stable outside this domain - up to $30 \; {\rm kpc}$ and beyond (i.e. $ Q_g > 1$ and $\Sigma_g < \Sigma_c$). However, when using the effective Toomre criterion (\ref{toomre-gmfdm}), one can see that DM extends the instability regime through a wider area across the disc between $1 - 13 \; {\rm kpc}$ (i.e. $Q_{ge} < 1$ and $\Sigma_g > \Sigma_{ce}$) while pushing the outer area towards the marginal stability line with an average value $Q_{ge} \sim 1.4$. This shows that when the gas is observed to be stable or marginally stable - using the standard threshold density $\Sigma_c$ - in areas of the disc where star formation is going on, it can be effectively unstable (i.e. super-critical) under the influence of DM: henceforth the measure of the density threshold should be $\Sigma_{ce}$, not $\Sigma_c$. On the other hand, MF (fig. \ref{Q-gasonlyMF} \& \ref{Sdens-gasonlyMF}) stabilises the gas disc throughout, even overcoming the effect of DM. 

\par
It is worth noting that in all the cases discussed above, the critical density follows the same profile as the gas distribution (fig. \ref{Sdens-gasonly} and  \ref{Sdens-gasonlyMF}). This might be an indication that the distribution of gas itself is determined in large part by the dynamics surrounding the stability of the disc (Elmegreen \& Hunter 2016). On the other hand, the stabilising properties of an axisymmetric MF should not be overemphasised since instabilities are favoured by non-axisymmetric MF at low shear. Particularly, in the regions of scarce SF in which we are interested, MHD instabilities are the main source of turbulence (Sellwood \& Balbus 1999).

\par
One cannot put too much weight to the results of this section as the dynamics of a galactic disc cannot be represented by a single gas sheet. It is more realistic to include the stellar component.

\subsection{A two-component disc }
\label{sec GS stability}

In an evolved stage of the disc, when star formation is underway, the stellar component changes the dynamics of the disc through the gravitational potential and heating (i.e. SNe feedback, stellar winds and generation of MF). We here consider that stars are coupled to gas and DM only through the gravitational potential that is now generated by DM and the total baryonic mass (gas + star) of surface density $\Sigma_{b} = \Sigma_{g} + \Sigma_s$, with $\Sigma_{s}$ the total stellar surface density. 

\par
Proceeding as in the case of gas alone (appendix B) and assuming that stars behave like a gas with a velocity dispersion $\sigma_{s}$, but not being affected by the galactic scale MF, one can show that the dispersion relation of a realistic disc made of stars, gas, DM and MF is

\begin{eqnarray}
\frac{2}{Q_{gdm}} \frac{\alpha_m K_g}{ 1 - (\omega / \kappa)^2 + (\alpha_m K_g)^2 } +
 \frac{2}{ Q_{sdm} } \frac{ K_{s} }{ 1 - (\omega / \kappa)^2 + K_{s}^2 } = 1   \;\;\; .
\label{dispsgh}
\end{eqnarray}

\noindent
Here $Q_{sdm} = (1 / \Gamma) Q_{s} $, with $Q_{s} = \kappa \sigma_{s} / \pi G \Sigma_{s}$ the criterion for the stellar component,  
$ Q_{gdm} = (\alpha_m / \Gamma ) Q_g $ and $\Gamma = 1 + \Sigma_{dm} / \Sigma_b$, with $\Sigma_{dm}$ the amount of halo DM embedded within the disc. We define the non-dimensional wavenumbers as $ K_g = k c_g / \kappa $ and $ K_{s} = k \sigma_{s} / \kappa $. This criterion is an extension of the criterion found by Jog \& Solomon (1984) with a contribution of DM and MF. 

\par
Figure \ref{Qparam-gas-star} shows the individual Q-parameters, $Q_g$ and $Q_s,$ for gas and star: both components are stable throughout the disc. However the stability of the stellar component ($Q_{sdm}$) is not much affected by DM whilst the gas component ($Q_{gdm}$) is made unstable throughout the disc from about 5 kpc outwards. On the other hand, MF has a strong stabilising effect on the gas component (fig. \ref{Qparam-gas-starMF}), reducing hence the effect of DM by expanding the stable area up to about 14 kpc from which the gas is unstable outwards (i.e. note that the epicyclic frequency in this situation is larger than in equation (\ref{toomreq})). It is significative that gas instability in the outskirts of the disc is driven by DM.

\par
We shall see in section \ref{sec effectiveparam} that the global stability of the disc is not rendered by the individual components but instead by an effective criterion that takes account of the gravitational coupling between star, gas and DM.

\section{The model}
\label{sec model}

\begin{figure}
\includegraphics[height=100mm, width=170mm,angle=0]{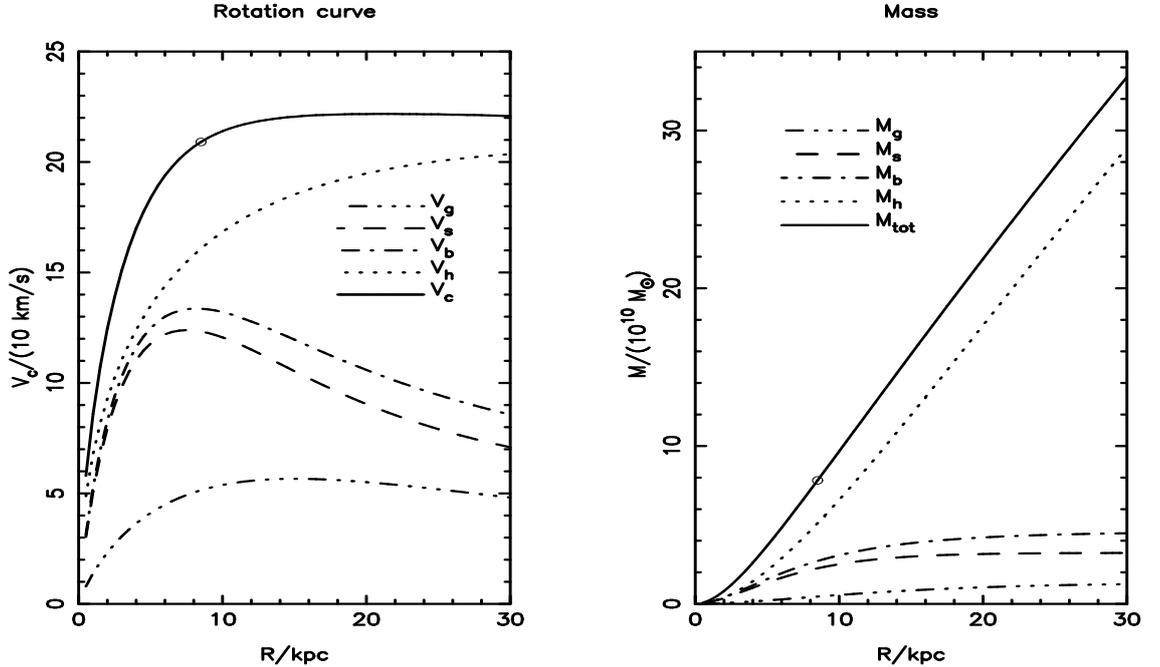}
\caption{Rotation curve (left panel) and mass model (right panel).}
\label{Vc-Mass}
\end{figure}

To illustrate the dynamics of the disc, we use the following ingredients that represent average propreties of a galactic disc like the Milky Way. We assume an exponential profile for the stellar and gas components

\begin{equation}
\Sigma_l = \Sigma_{0l} \exp\left(- \frac{R}{R_l}\right) 
\end{equation}

\noindent
with $l = g \; {\rm or} \; s$ (for gas or star). The radial scale length $R_s = 3 \pm 0.5 \; {\rm kpc }$ and $R_g = 2 R_s$: this ensures that the gas component extends way beyond the luminous stellar component as observed in galactic discs. The values of $\Sigma_{0l}$ are normalised at the solar position $R_0$, with $\Sigma_{g} (R_0) = 13 \; {\rm M_{\odot} pc^{-2}} $ and $\Sigma_{s} (R_0) = 37 \; {\rm M_{\odot} pc^{-2}}$, including stellar remnants (BT08) - we choose $\Sigma_{b} (R_0) = 50 \; {\rm M_{\odot} pc^{-2}}$ for the baryonic surface density.
In general $\Sigma_{b} = 35-58 \; {\rm M_\odot pc^{-2}}$ (Weber \& de Boer, 2009). However McMillam (2011) suggests much higher mean values $\Sigma_{b} = 62.0 \pm 7.6 \; {\rm M_\odot pc^{-2}}$.
Comparison between total and visible baryonic matter leaves a discrepancy of about $20-30 \; {\rm M_\odot pc^{-2}}$ of unidentified matter which in the present study could be identified to a contribution from a spherical dark halo.

\par
Analysis of the radial velocity and photometric survey of local K dwarfs within $z = 1.1 \; {\rm kpc}$ (Kuijken \& Gilmore 1989) and modeling of the rotation curve led to values of the total surface density $\Sigma_{1.1} = 71 \pm 6 \; {\rm M_\odot pc^{-2}}$ and the baryonic surface density $\Sigma_{b} = 48 \pm 9 \; {\rm M_\odot pc^{-2}}$ (Kuijken \& Gilmore, 1989, 1991). These authors concluded that there is no dark matter associated with the disc.
  
\par
Holmberg \& Flynn (2004) analysing {\it Hipparcos} K giants found $\Sigma_{1.1} = 74 \pm 6 \; {\rm M_\odot pc^{-2}}$, $\Sigma_{0.8} = 65 \pm 6 \; {\rm M_\odot pc^{-2}}$, $\Sigma_{0.35} = 41 \; {\rm M_\odot pc^{-2}}$ and $\Sigma_b \sim 52.5 \; {\rm M_\odot \; pc^{-2}}$. Bienaym\'e et al. (2005) found 
$\Sigma_{1.1} = 68 \pm 11 \; {\rm M_\odot pc^{-2}}$. Catena \& Ullio (2010) found the mean values for an NFW profile $\Sigma_{b} = 46.24 \pm 5.38 \; {\rm M_\odot pc^{-2}}$ and $\Sigma_{1.1} = 72.13 \pm 4.18 \; {\rm M_\odot pc^{-2}}$.

\par
The rotation velocity of each exponential disc component  is (Freeman 1970, Chamcham \& Tayler 1995, BT08)

\begin{equation}
V_{l}^{2} (R) = 4 \pi G \Sigma_{0l} R_l Y_{l}^{2} \Bigl\{ I_0 (Y_l) K_0 (Y_l) - I_1 (Y_l) K_1 (Y_l) \Bigr\} \; , {\rm where} \; 
Y_l = \frac{R}{2 R_l} \; ,
\label{Vrotexponential}
\end{equation}

\noindent
$ I_n \; {\rm and} \; K_n $ are the modified Bessel functions of first and second kinds. The associated epicyclic frequency

\begin{eqnarray}
\kappa_{l}^{2} (R) & = & \frac{4 \pi G \Sigma_{0l}}{R_l} \biggl\{ I_0 (Y_l) K_{0} (Y_l) - \frac{1}{2} I_1 (Y_l) K_1 (Y_l) \biggr.   
\nonumber   \\                    
				   & + & \frac{Y_l}{2} \biggl. \biggl(I_1 (Y_l) K_0 (Y_l) - I_0(Y_l) K_1(Y_l) \biggr) \biggr\}  \;\; .
\label{epicyclexponential}
\end{eqnarray}

\par
Observations show that the velocity dispersion of clouds in all galaxies is nearly isotropic (Binney \& Merrifield 1998), with $\sigma_{\rm H_I} = 11 \; {\rm km s^{-1}}$ (Leroy et al., 2008) and $\sigma_{\rm H_2} = 6 \; {\rm km s^{-1}}$. We treat the ISM as a single component with $\Sigma_g = \Sigma_{\rm H_I} + \Sigma_{\rm H_2} $ and $c_g = 6 \; {\rm km s^{-1}}$. Silk (1997) argues that the velocity dispersion is predicted to remain roughly constant in self-regulated regions. 
Although the star velocity dispersion seems to follow an exponential profile, $ \sigma_s \propto \exp( - R/R_{s}) $ (Lewis \& Freeman 1989 ), we shall use a constant value $\sigma_s = 25 - 45 \; {\rm km s^{-1}}$ throughout the disc. 

\par
We choose a NFW density profile for the DM halo (NFW 1997), normalised at $R_0$

\begin{equation}
\rho_{h} (r) = \rho_{dm \odot} \; u_0 (1 + u_0)^2 \left( \frac{r}{R_c} \right)^{-1} \frac{1}{\left(1 + r/R_c\right)^{2} } \;\; ; \; r^2 = R^2 + z^2
\end{equation}

\noindent
with the associated rotation curve at $z = 0$

\begin{equation}
V^2_{h} (R) = 4 \pi G \rho_{dm \odot} \; u_0 (1 + u_0)^2 R_c^2 \frac{1}{u} \left\{ \ln (1 + u) - \frac{u}{1 + u} \right\} 
\label{RC-NFW}
\end{equation}

\noindent
and the epicyclic frequency

\begin{equation}
\kappa^2_{h} (R) = 4 \pi G \rho_{dm \odot} \; u_0 (1 + u_0)^2 \frac{1}{u^3} \left\{ \ln (1 + u) - \frac{u}{(1 + u)^2} \right\} 
\label{epicycl-NFW}
\end{equation}

\noindent
with $u = R/R_c$, $u_0 = R_0/R_c$. We use $R_c = 20$ kpc which was deduced by Weber \& de Boer (2010) using an NFW profile. The disc rotation curve is

\[ V^2_{c} = V^2_{g} + V^2_{s} + V^2_{h}  \]

\noindent
and the epicyclic frequency
\[\kappa^2 = \kappa^2_{g} + \kappa^2_{s} + \kappa^2_{h} \; . \]

We adopt the value of the local DM density from Hooper (2017): in discussing the contribution of DM in the bulge of the Galaxy he uses a local density $\rho_{dm \odot} = 0.4 \; {\rm GeV cm^{-3}}$ and suggests that the results of Portail et al. (2015) cover a range of values $\rho_{dm \odot} = 0.156 - 0.504 \; {\rm GeV cm^{-3}}$. 

\par
The local dark matter density constrained from different observations such as microlensing, the total projected mass density, peak to trough variations in the rotation curve (i.e. the flatness constraint) lies within the range 
$\rho_{dm\odot} \cong 4 \; 10^{-25} \; {\rm g \; cm^{-3}} - 13 \; 10^{-25} \; {\rm g \; cm^{-3}}$ - i.e. 
$0.22 - 0.73 \; {\rm GeV cm^{-3}}$ 
(Gates, Gyuk \& Turner 1996, Amsler et al. 2008). Constraints on N-Body simulations run by Ling et al. (2011) provide a value $ \rho_{dm\odot} = 0.37 \; {\rm GeV cm^{-3}} $. 

\par
The halo parameters are also constrained by the total mass within a radius R. Our model predicts
$ M_{50} = 5.37 \; 10^{11} \; {\rm M_{\odot}}$, $ M_{60} = 6.27 \; 10^{11} \; {\rm M_{\odot}}$ and $ M_{100} = 9.21 \; 10^{11} \; {\rm M_{\odot}}$, respectively for $R = 50, 60$ and $100 \; {\rm kpc}$. 
McMillan (2011) deduced the average values and standard deviations $ M_{50} = 5.1 \pm 0.4 \; 10^{11} \; {\rm M_{\odot}}$, $ M_{60} = 5.9 \pm 0.5 \; 10^{11} \; {\rm M_{\odot}}$ and $ M_{100} = 8.4 \pm 0.9 \; 10^{11} \; {\rm M_{\odot}}$ with a virial mass $ M_{v} = 1.26 \pm 0.24 \; 10^{12} \; {\rm M_{\odot}}$. 

\par
Various tracers have been used to determine the total mass of the Galaxy, such as hypervelocity stars (Fragione \& Loeb 2017), kinematics of distant halo tracer stars (Xue et al. 2008), satellite galaxies and vertical scale height of the gas distribution of the Galactic disc. Battaglia et al. (2006) found
$ M_{tot} = 0.3-2.5 \; 10^{12} \; {\rm M_{\odot}}$ while Wilkinson \& Evans (1999) derived
$ M_{tot} = 1.9^{+3.6}_{-1.7} \; 10^{12} \; {\rm M_{\odot}}$ but they found that the mass within 50 kpc, 
$ M_{50} = 5.4^{+0.2}_{-3.2} \; 10^{11} \; {\rm M_{\odot}}$, is more robustly determined. 
Catena \& Ullio (2010), using an MCMC method deduced for an NFW profile $ M_{50} = 5.35 \pm 0.24 \; 10^{11} \; {\rm M_{\odot}} $ and $ M_{100} = 8.56 \pm 0.53 \; 10^{11} \; {\rm M_{\odot}} $ with a virial mass
$ M_v = 1.49 \pm 0.17 \; 10^{12} \; {\rm M_{\odot}} $. 

\par
Figure (\ref{Vc-Mass}) shows our model rotation curve and the associated mass model, with $V_c(R_0) = 210 \; {\rm km s^{-1}}$ at the solar neighbourhood,  then flattening at around $220 \; {\rm km s^{-1}}$ from the solar position outwards. Correspondingly, considering all the above constraints, our model predicts a DM surface density
$\Sigma_{dm} (R_0) = 26.4 \; {\rm M_\odot pc^{-2}}$, then steadily decreasing to about $13.3 \; {\rm M_\odot pc^{-2}}$ at 30 kpc.

\begin{figure}
\includegraphics[height=100mm, width=170mm,angle=0]{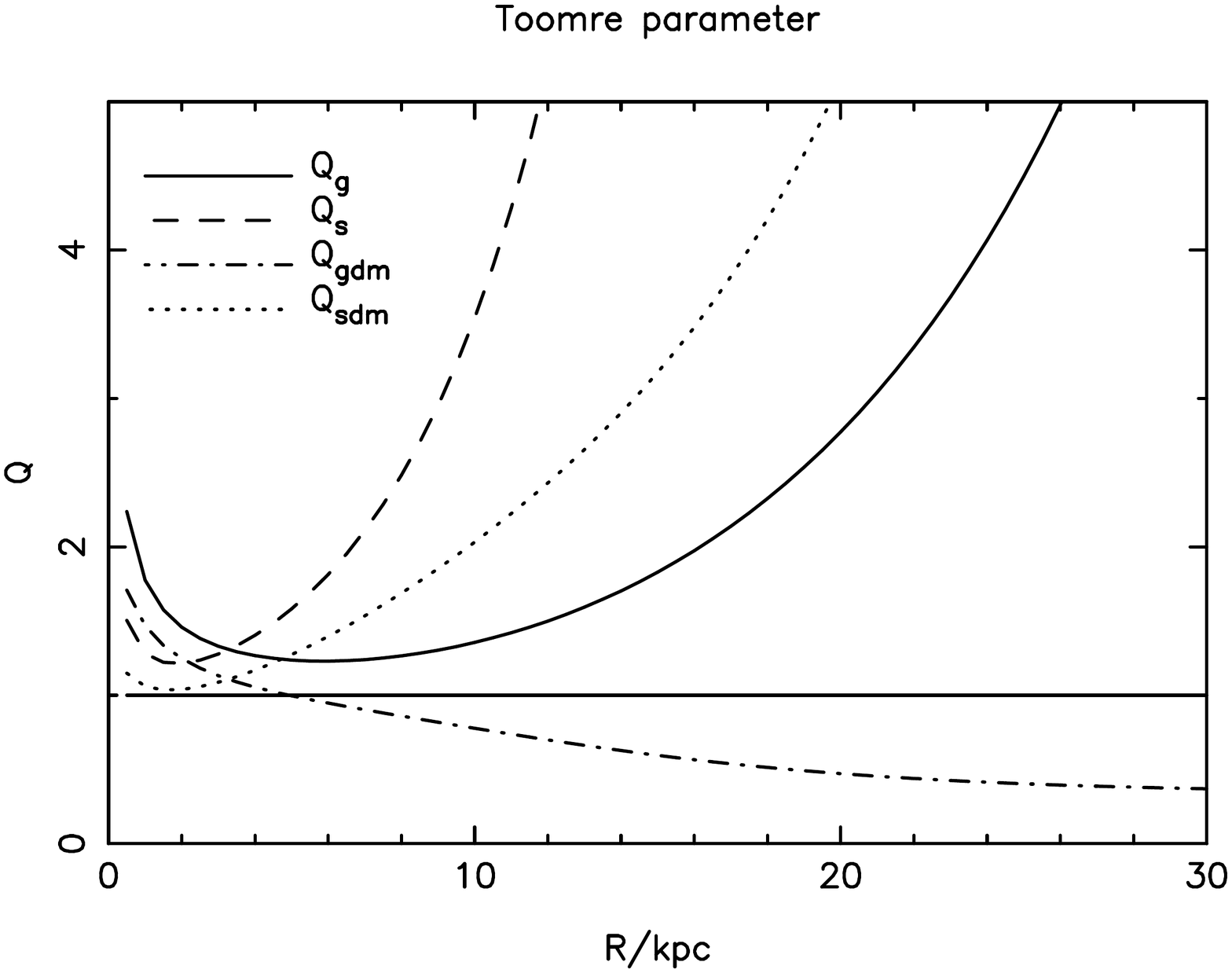}
\caption{Q-parameter for star (dash) and gas (continuous) treated as separate fluids within the disc. The effect of DM is shown on gas (dash-dot) and star (dot).}
\label{Qparam-gas-star}
%
\includegraphics[height=100mm, width=170mm,angle=0]{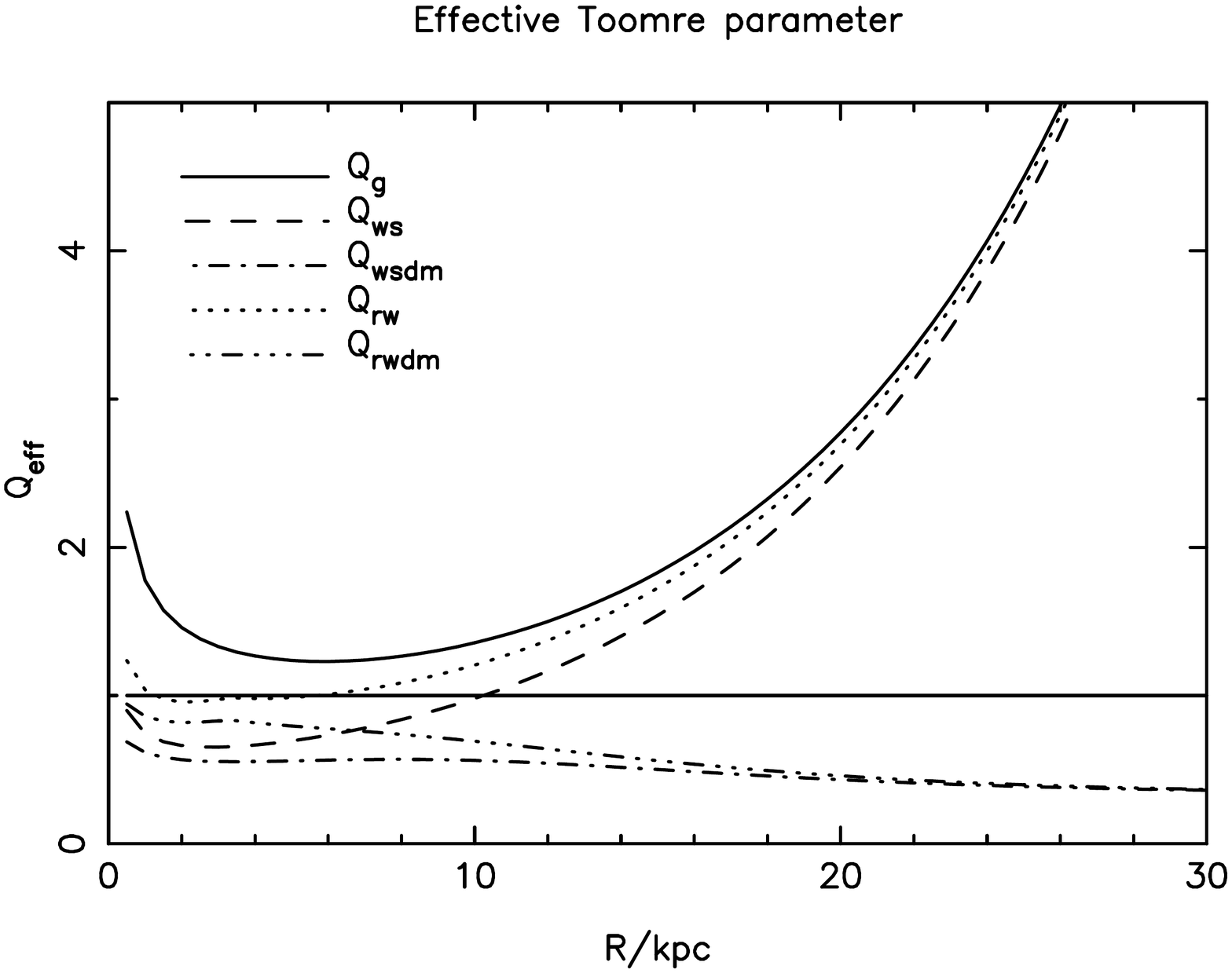}
\caption{Effective Q-parameter for star+gas in Wang \& Silk (dash) and Romeo \& Wiegert (dot) approximations. The effect of DM is shown on WS (dash-dot) and RW (dash-dot-dot-dot). The Q-parameter for gas alone is shown for comparison.}
\label{Qeff-param}
\end{figure}

\section{ The effective Q parameter vs sub-critical star formation }
\label{sec effectiveparam}

Apart from the simplifying assumption used by many authors (Kennicutt 1989, Martin \& Kennicutt 2001, Hunter et al. 1998, Schaye 2004) where the contribution of the stars to the stability of the disc is introduced as a correction to the criterion for gas alone, 

\begin{equation}
Q_{ef} = \alpha Q_g \; , 
\label{Qeff}
\end{equation}

\noindent
there is no easy analytic solution for  developing an effective criterion, $ Q_{ef} $, for a multi-fluid system as a function of the criterion of each of the system components. 

\par
Kennicutt (1989) used a value $ \alpha = 0.7 \pm 0.2 $ throughout the disc, assuming a constant gas velocity dispersion $c_g = 6 \; {\rm km/s}$ across all galaxies of his sample. Other authors (Pisano et al. 2000) used $c_g = 10 \; {\rm km/s}$ to find $\alpha = 0.3$ (instead of 0.5 if they used $c_g = 6 \; {\rm km/s}$). However, Martin \& Kennicutt (2001) showed that the azimuthally averaged values of the gas surface density and the SFR in their sample can lead to errors in $\alpha(R_{ \rm H\,{\sc II} })$ as large as a factor of 2 when discs are highly nonaxisymmetric. 

\par
On the other hand, it is not straightforward to show that $Q_{ef}$ is directly linked to SF or that the marginal value $Q_{ef} = 1$ is linked to a gas critical density (Elmegreen 1995, 2011). However we can imagine that if the multi-fluid system is unstable, in an active SF region, the gas component for instance will be unstable to allow for SF to occur and hence we can define an effective gas critical density $\Sigma_{cef}$ for $Q_{ef} = 1$. In any case a multi-fluid system is always less stable than each of its single components, which means that it is more likely that the gas effective critical density is lower than $\Sigma_c$ for gas alone, and possibly leading to the case: $\Sigma_{cef} < \Sigma_g < \Sigma_c$ where SF is effectively super-critical whilst it appears sub-critical if scrutinised using the simple Toomre criterion (\ref{toomreq}) - or the critical density $\Sigma_c$ - as a reference. Also, during the process of formation of the disc, it is important that the stellar component remains stable and henceforth only the gas component is concerned with the instability for the ongoing SF. 
In the case of Kennicutt's parametrization, the effective critical density is $ \Sigma_{cK} = \alpha \Sigma_c $, where the parameter $\alpha$ takes account of the presence of stars. 

\begin{figure}
\includegraphics[height=100mm, width=170mm,angle=0]{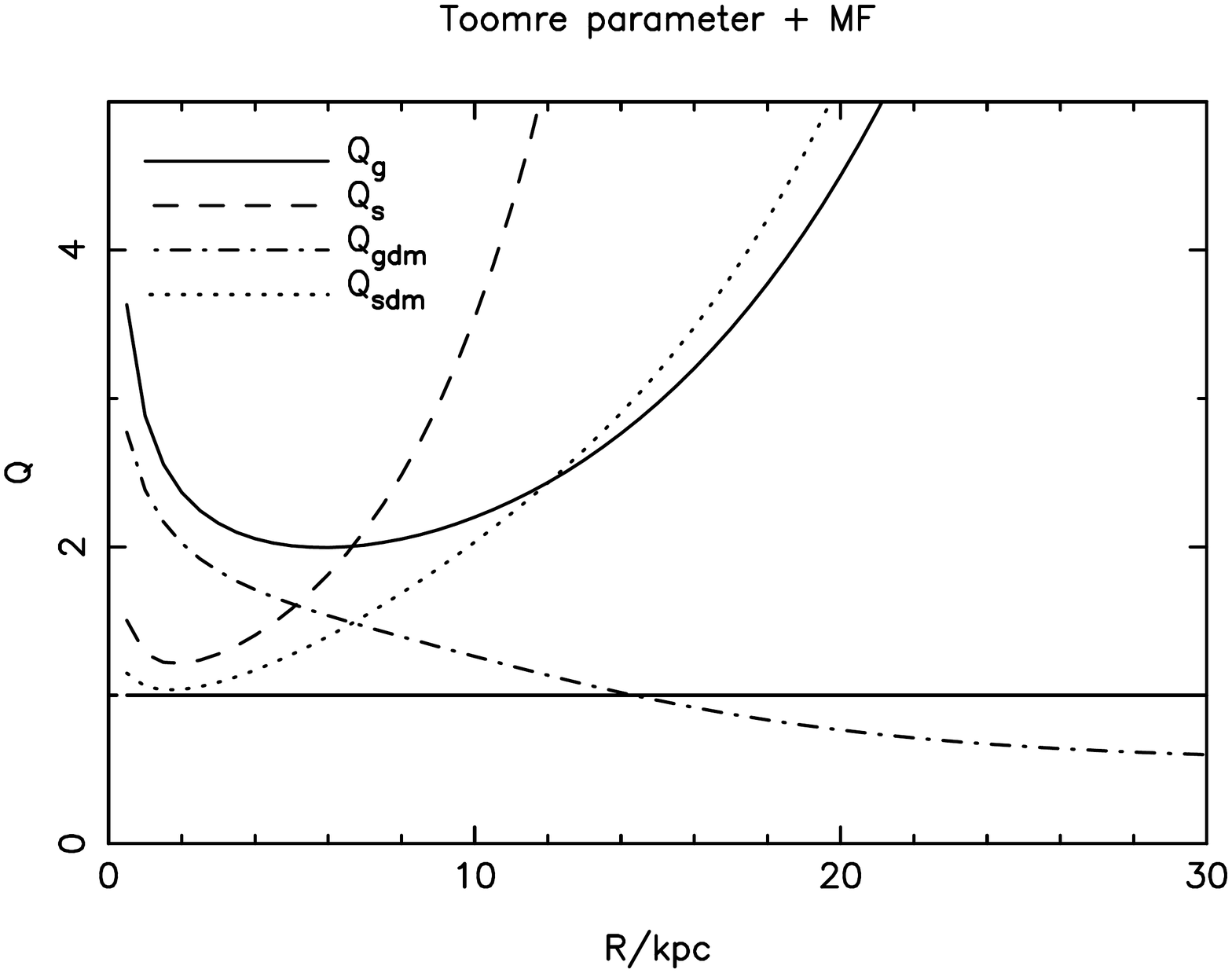}
\caption{Identical to figure \ref{Qparam-gas-star} but with the effect of MF.}
\label{Qparam-gas-starMF}
%
\includegraphics[height=100mm, width=170mm,angle=0]{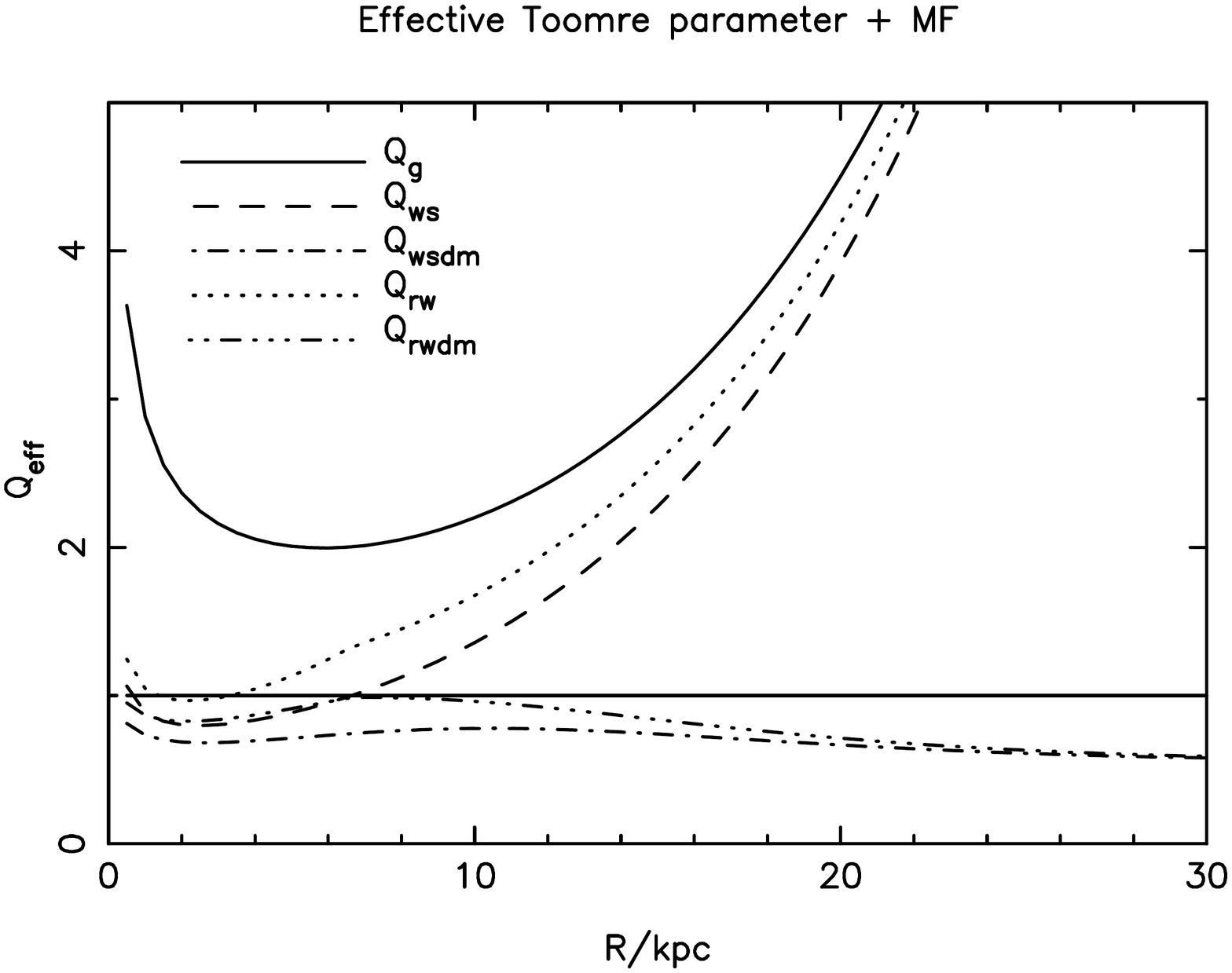}
\caption{Identical to figure \ref{Qeff-param} but with the effect of MF.}
\label{Qeff-paramMF}
\end{figure}

\par
In the presence of stars the simplest criterion for the (gas + stars) fluid occurs when 
$ \sigma_s = c_t = \sigma $. This situation can be found before the stars are heated - at early phases of star formation - when they are still embedded within the remnants of the clouds out of which they formed (eg. SF within the pillars); in this case the effective criterion for the total baryons is

\begin{equation}
Q_{b} = \frac{\kappa \sigma }{ \pi G \Sigma_b } = \alpha_m \mu_g Q_g  \; ,
\label{cgcs}
\end{equation}

\noindent
where $\mu_g$ is the gas fraction. The Kennicutt parameter $\alpha$ is simply $\alpha_m \mu_g$ and the 
effective critical density is 

\begin{equation}
\Sigma_{cb} = \alpha_m \Sigma_c - \Sigma_s \; ,
\end{equation} 

\noindent
as long as the stars are stable (i.e. $Q_s > 1$) - a condition that can easily be met at early stages of SF.
This shows that SF can be very efficient within clouds at the early phases - within gas rich regions - before feedback becomes important as stars evolve, hence reducing the process of SF. Eventually, the star velocity dispersion and surface density will increase with time and this approximation will cease to be valid. What is interesting is that the onset of early star formation reduces the threshold, allowing smaller quantities of gas to still form stars.

\par
In presence of dark matter the effective criterion (\ref{cgcs}) becomes

\begin{equation}
Q_{bdm} = \frac{\kappa \sigma }{\Gamma \pi G \Sigma_b } = \frac{\alpha_m \mu_g}{\Gamma} Q_g  \; ,
\label{cgcsdm}
\end{equation}

\noindent
the Kennicutt parameter $\alpha_{bdm} = \alpha_m \mu_g / \Gamma$ and the effective threshold density becomes 

\begin{equation}
\Sigma_{cbdm} = \alpha_m \Sigma_c - \Sigma_s - \Sigma_{dm} \; ,
\end{equation}

\noindent
showing that DM lowers even further the threshold density for star formation to occur.

\par 
A simplified formulation of the effective stability criterion has been suggested by Wang \& Silk (1994). Assuming that the two fluids react independently to the perturbation this leads to the analytic formulation 

\begin{equation}
 \frac{1}{Q_{ws}} = \frac{1}{Q_g^{'}} + \frac{1}{Q_s^{'}}    
\label{silk}    
\end{equation}

\noindent
This shows that the system is less stable than any one of its components taken separately\footnote{We use the notations $Q_g^{'}$ and $Q_s^{'}$ as we will be replacing these quantities either by their standard values $Q_g$ and $Q_s$ or their effective values including the effects of MF or both MF+DM as in (\ref{dispsgh}).} . In this case the Kennicutt parameter $\alpha$, including MF, is

\begin{equation}
\alpha_{ws} = \alpha_m \left( 1 + \alpha_m \eta \frac{\Sigma_s}{\Sigma_g} \right)^{-1} \; ,
\label{alfaWS}
\end{equation}

\[ \;\;  {\rm with} \; \eta = \frac{c_g}{\sigma_s} 
\; , \;\; \frac{Q_g}{Q_s} = \eta \frac{\Sigma_s}{\Sigma_g} \; \]

\noindent
and the corresponding effective critical density 

\begin{equation}
\Sigma_{cws} = (1 - 1/Q_s) \Sigma_c = \alpha_m ( \Sigma_c - \eta \Sigma_s ) \; .
\label{scritWS}
\end{equation}
 
\noindent
In addition the presence of DM will reduce the Kennicutt parameter $\alpha$ with the quantity $\Gamma$ such that

\begin{equation}
\alpha_{wsdm} = \frac{\alpha_m}{\Gamma} \left( 1 + \alpha_m \eta \frac{\Sigma_s}{\Sigma_g} \right)^{-1} \; ,
\label{alfaWSDM}
\end{equation}

\noindent
and the new effective critical density is given in its general form by (\ref{ScritRW}).

\begin{figure}
\includegraphics[height=100mm, width=170mm,angle=0]{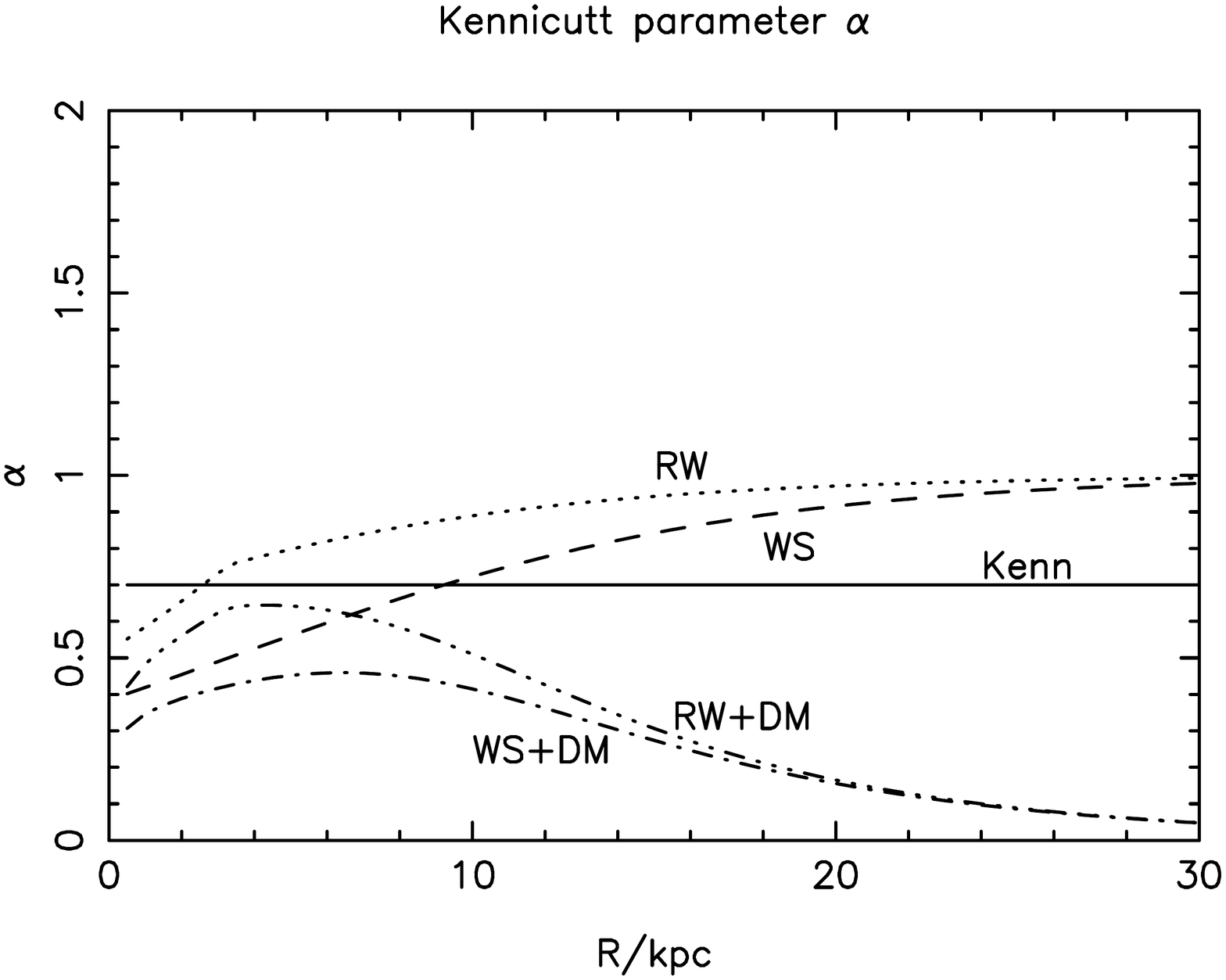}
\caption{Kennicutt $\alpha$ parameter in the cases of WS and RW with and without the effect of DM.}
\label{alfaKen}
%
\includegraphics[height=100mm, width=170mm,angle=0]{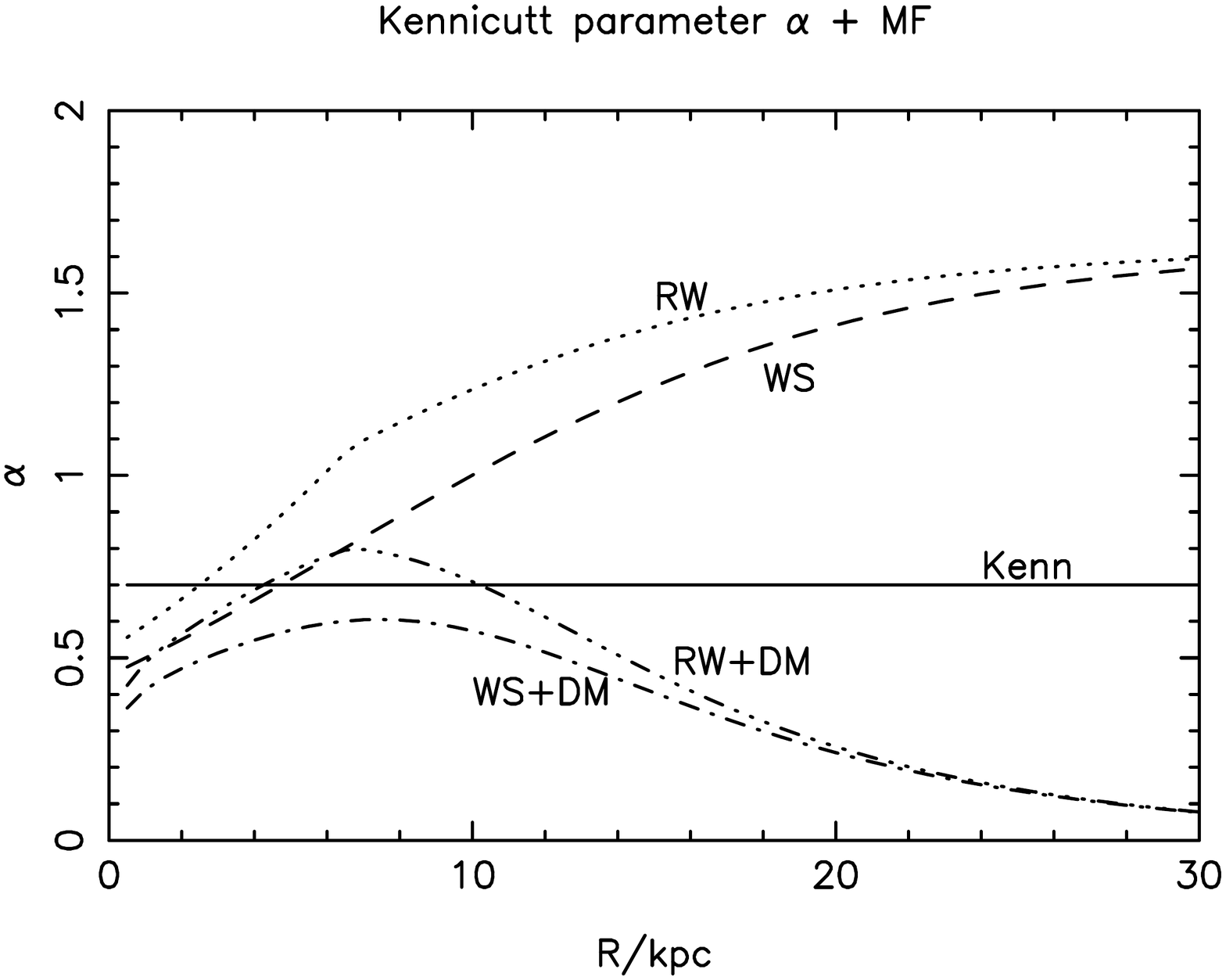}
\caption{Kennicutt $\alpha$ parameter in the cases of WS and RW with and without the effect of DM, but with the effect MF.}
\label{alfaKenMF}
\end{figure}
%
%

\par
However, Romeo and Wiegert (2011; their figure 2) argue that the effective parameter deduced by Wand and Silk (1994) carries large uncertainties. They suggest a more accurate approximation for a thin disc (i.e. without the effect of the thickness)

\begin{equation}
\frac{1}{Q_{rw}} = \frac{W_s}{Q_s^{'}} + \frac{W_g}{Q_g^{'}}  \left\{ \begin{array}
          {r@{\quad:\quad}l}
           W_g =1, \;  W_s = W & Q_s^{'}  \ge Q_g^{'}   \vspace{0.5cm}  \\   W_g = W, \; W_s = 1 &   Q_g^{'}  \ge Q_s^{'}    
                \end{array}    \right.
\label{critRW}
\end{equation}

\noindent
with

\[ W = \frac{2 \sigma_s \sigma_g}{\sigma_s^2 + \sigma_g^2} \; , \sigma_g = c_t = \alpha_m c_g  \]

\noindent
one obtains
%

\begin{equation}
\alpha_{rw} = \alpha_m \left( W_g + W_s \alpha_m \eta \frac{\Sigma_s}{\Sigma_g} \right)^{-1} \; ,
\label{alfaRW}
\end{equation}
\noindent

and

\begin{equation}
\Sigma_{crw} = \frac{\alpha_m}{W_g} ( \Sigma_c - W_s \eta \Sigma_s ) \; .
\label{scritRW}
\end{equation}
%

\noindent
They compared the stability threshold determined numerically by Bertini and Romeo (1988) to the WS criterion and found that the error can be up to -50\%, hence concluding that the WS approximation underestimates the effective Q-parameter systematically.

\par
Consideration of the role of dark matter leads to

\begin{equation}
\alpha_{rwdm} = \frac{\alpha_m}{\Gamma} \left( W_g + W_s \alpha_m \eta \frac{\Sigma_s}{\Sigma_g} \right)^{-1}
\end{equation}

\noindent
and the effective critical density

\begin{equation}
\Sigma_{crwdm} = \frac{W_g \Sigma + \alpha_m \eta W_s \Sigma_s - \alpha_m \Sigma_c}{2 W_g} \left\{ -1 +
\sqrt{ 1 - \frac{4 \alpha_m W_g \Sigma_s (\eta W_s \Sigma - \Sigma_c )}{ [ W_g \Sigma + \alpha_m \eta W_s \Sigma_s - \alpha_m \Sigma_c ]^2 } } \right\} \; ,
\label{ScritRW}
\end{equation}

\noindent
with $ \Sigma =  \Sigma_s + \Sigma_{dm} $. We recover the effective critical surface density, $\Sigma_{cws}$, for the case of Wang \& Silk by setting $W_g = W_s = 1$.

\begin{figure}
\includegraphics[height=100mm, width=170mm,angle=0]{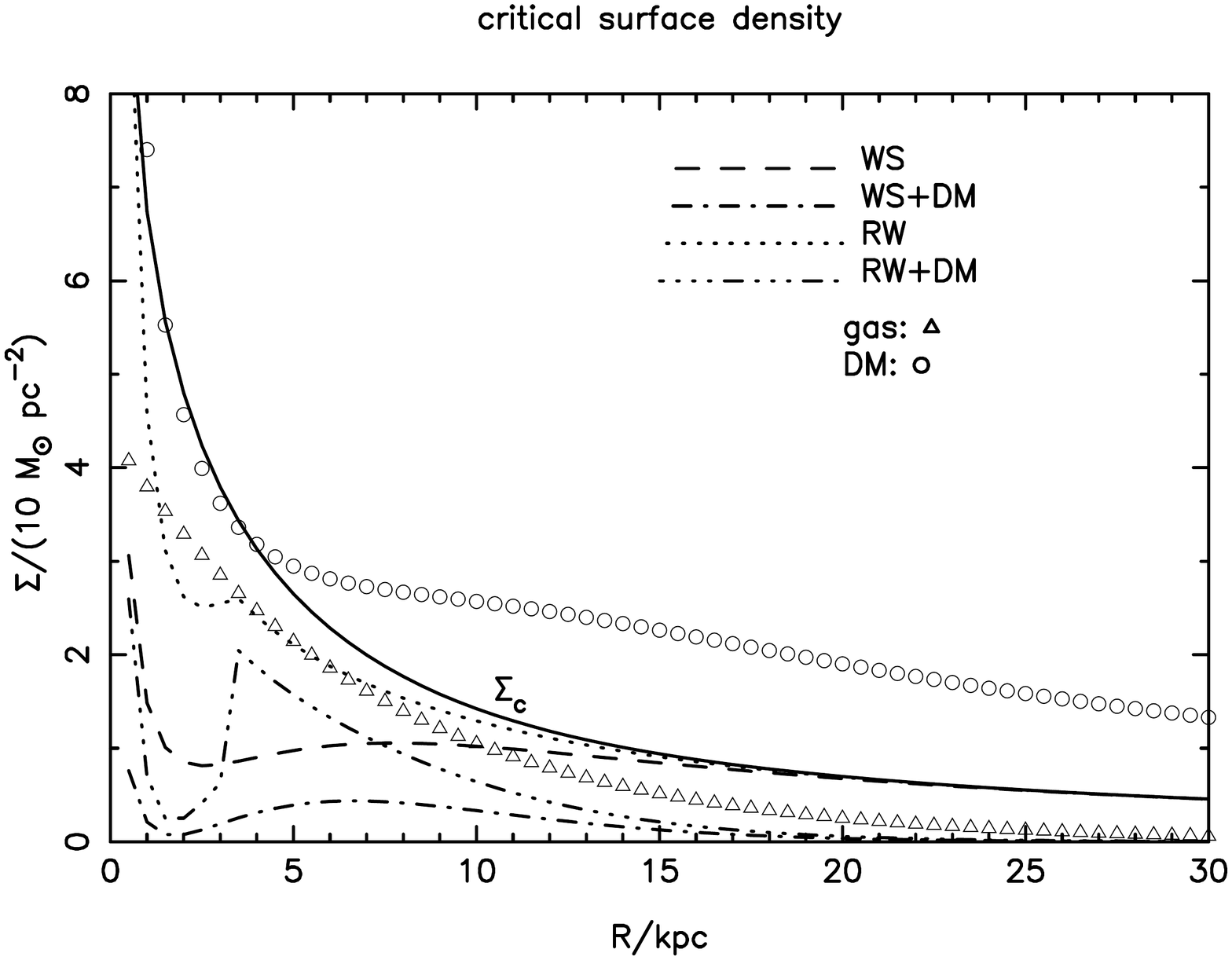}
\caption{The effective critical surface density with the effect of DM.}
\label{Sdens-Scrit}
%
\includegraphics[height=100mm, width=170mm,angle=0]{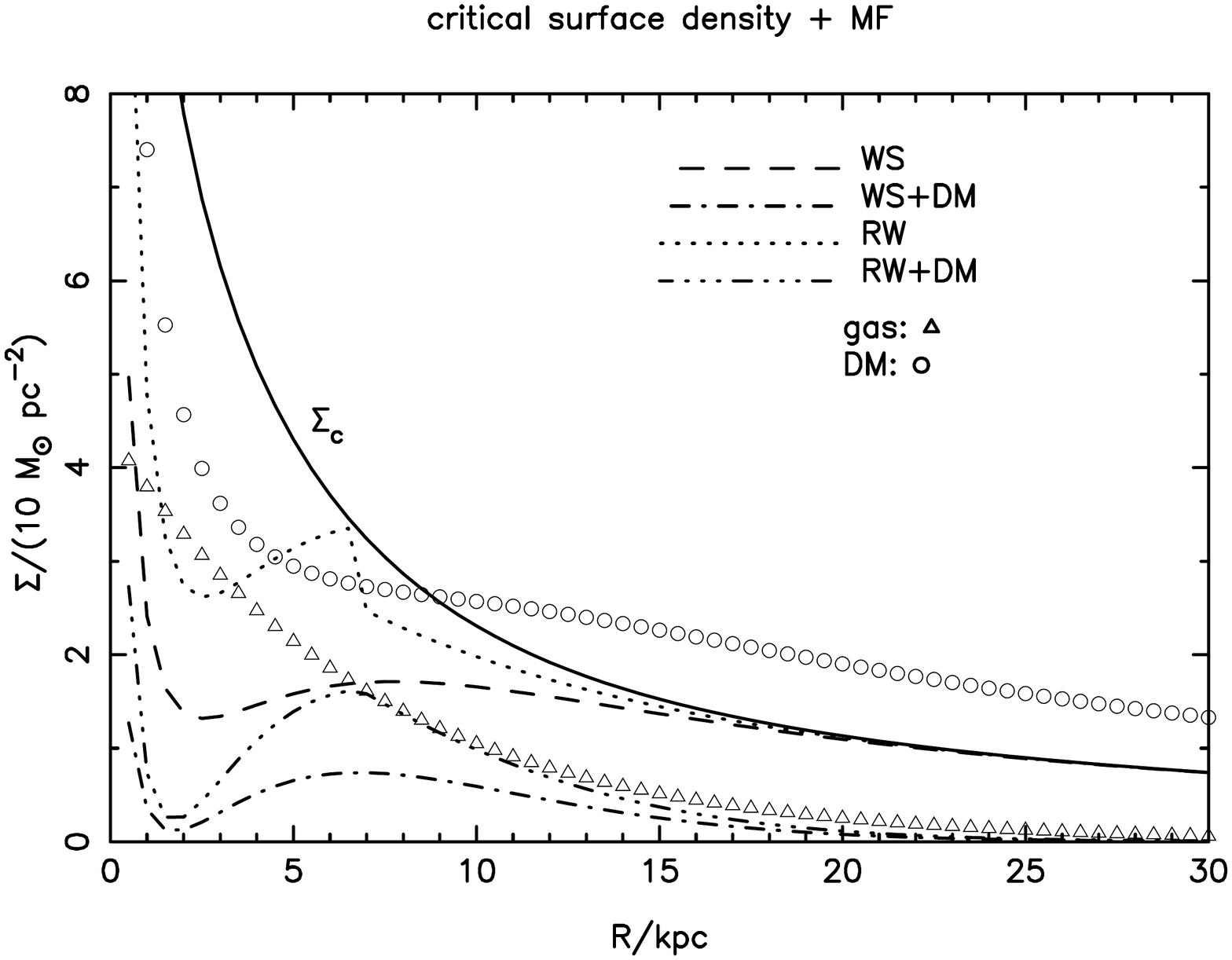}
\caption{Identical to figure \ref{Sdens-Scrit} but with the effect of MF.}
\label{Sdens-ScritMF}
\end{figure}

\par
It becomes clear from the above discussion that sub-criticality is a matter of definition of the effective critical density, based on the approximations used to work out the effective Toomre stability criterion and consideration of the 'ingredients' that play into regulating gas dynamics and hence the process of SF. If this be $\Sigma_c$, then discs can be seen as sub-critical (while instabilities are driven by the unseen/neglected DM). Based on this preliminary study we argue that the real threshold for star formation should be measured with the effective critical density $\Sigma_{cef}$ related to the effective criterion $Q_{ef}$ which is not easy to determine without assumptions on the processes regulating the dynamics of the disc. Still the occurrence of SF at a much lower gas density than expected in some parts of the galactic disc is the imprint of DM that 'boosts up' the self-gravity of the gas layers and drives them towards instability at much lower thresholds than expected when using the standard Toomre criterion.

%

\section{Discussion}
\label{sec discuss}

Figure \ref{Qeff-param} shows that while both the gas and star components are stable (fig. \ref{Qparam-gas-star}) the disc can be unstable (Wang \& Silk approximation: WS) or marginally stable (Romero \& Wiegert approximation: RW) for some part of the disc: instability spreads through an area between 1.5-6 kpc (RW) and between 0.5-10.5 kpc (WS). However, the presence of DM makes the disc unstable inside-out as in LSB galaxies - this is also particularly interesting for the outskirts of disc galaxies which exhibit 'peculiar' SF where the disc seems stable when DM effects are neglected. 

\par
In the present picture, the presence of DM as a driver of SF seems to offer an explanation to the occurence of SF where the gas is observed to be stable when scrutinised using the standard criterion (\ref{toomreq}): eg. the existence of knots at the outskirts of galaxies, where gas seems stable, might reveal the double signature of the workings of DM as well as its lumpy structure. In relation to this, figure (\ref{alfaKen}) shows that using a constant value, throughout the disc, of the Kennicutt parameter $\alpha$ does not account for the dynamics of the disc in the same way as WS or RW do (equs. (\ref{alfaWS}) and (\ref{alfaRW})): it is possible that the wide uncertainty in the Kennicutt's value and the very low values discussed by some authors reflect the existence of this overlooked dynamics, particularly the role of DM which drives $\alpha$ to much lower values than 0.7. Note that the claim by Romeo \& Wiegert that their approximation is more accurate than that of Wand \& Silk does not have strong grounds within this analysis since both approximations follow the same trend and converge from about 10 kpc throughout the outskirts. 

\par
The presence of MF has a significant stabilising effect on the gas component which is the driver of instability in this model (figs. \ref{Qparam-gas-star}, \ref{Qparam-gas-starMF}). Comparing figure (\ref{Qeff-paramMF}) to (\ref{Qeff-param}) shows that the domain of instability is reduced to 1.5-3.5 kpc (RW) and to 1-6.5 (WS): i.e. when competing with DM, MF push the inner disc toward the line of stability but DM remains dominant outwards keeping the disc in the instability regime. This is an important result for galaxies such as LSBs and even our Galaxy, where star formation has been evolving slowly for the last few Gyrs, as well as for the outskirts of dIrrs where the gas is observed to be stable but SF seems very efficient. One can infer from this that SF efficiency is not particularly related to high gas abundance: i.e. SF can be very efficient in gas poor regions if local dynamical processes such as turbulence, coupling of baryons to DM combine to strengthen the disc self-gravity. Note that in all cases instability is driven by gas while the stellar component remains stable or marginally stable within a narrow inner disc area.

\par
Correspondingly, looking at the surface density distribution of gas within the disc, figure (\ref{Sdens-Scrit}) shows that the gas is stable ($\Sigma_g < \Sigma_c$) throughout when its distribution is compared to the standard threshold density derived from the Toomre criterion (\ref{toomreq}). But it is super-critical ($\Sigma_g > \Sigma_{cws}$) up to about 10.5 kpc when checked against the WS effective threshold density (\ref{scritWS}) and within 1.5-6 kpc ($\Sigma_g > \Sigma_{crw}$) if checked against the RW effective threshold density (\ref{scritRW}). However, the disc is super-critical throughout in both cases when DM is considered (\ref{ScritRW}). Note that both WS and RW approximations converge from about 15 kpc outwards. However, MF does not change this trend but it reduces the instability domain (fig. \ref{Sdens-ScritMF}).

\par
A remarkable feature in all the cases shown above is that the critical density and the gas distribution follow the same profile as the DM distribution (figs. \ref{Sdens-Scrit} and  \ref{Sdens-ScritMF}) and that the gas everywhere in the disc lies near the threshold for SF, i.e. the distribution of gas itself maybe determined in large part by the stability condition that is driven by DM. This is in accordance with observations which show that there is a general property of the gas discs in late-type galaxies to lie near the line of gravitational stability. It seems, from the present work, that these profiles are imprinted by the DM surface density profile: i.e. the presence of DM turns any amount of gas into stars - or at least coalesces any amount of gas into a compact structure. If this result is confirmed by a more detailed study and observations it will shed lights on the process of SF and stellar evolution with some interesting consequences on the IMF.

\par
The other interesting result of this work is the existence of an effective threshold density, $\Sigma_{cef}$ associated to the condition $Q_{ef} = 1$, for a realistic disc, and that the disc instability condition $Q_{ef} < 1$ is correlated to $\Sigma_g >\Sigma_{cef}$. These results can be improved by treating stars and DM as non-collisional fluids, with a specific equation of state for DM. 


\section*{Appendix A: Unperturbed equations }
\label{sec Gunperturb1}

The MHD equation for a sheet of gas of surface density $\Sigma_g$, moving with a velocity
$\vec{v}_{0} \equiv ( v_{0R}, v_{0\varphi}, v_{0z} )$, in equilibrium under the influence of a gravitational potential $\Phi_{0}$ and MF $\vec{B}_{0} \equiv ( B_{0R}, B_{0\varphi}, B_{0z} ) $ is the following 

\begin{eqnarray}
{\partial \vec{v}_{0} \over \partial t} + (\vec{v}_{0} \cdot \nabla) 
\vec{v}_{0} & = &
- {1 \over \Sigma_{g}} \nabla P_{0} - \nabla \Phi_{0} - 
{1 \over \Sigma_{g}} \nabla ( {\vec{B}_{0}^{2} \over 8 \pi} )   \nonumber   \\
            &   & 
+ {1 \over 4 \pi \Sigma_{g}} (\vec{B}_{0} \cdot \nabla) \vec{B}_{0} \;\; .
\label{vunpert} 
\end{eqnarray}

\noindent
The original equation is expressed in term of the volume density $\rho_g$, but we here replace it with the surface density $\Sigma_g$ because we are dealing with a two-dimensional thin disc and we are not considering the effect of its thickness. The passage to equation (\ref{vunpert}) is obtained by rescaling the gas pressure and the magnetic pressure terms with a scale-height $H$ such that $\Sigma_g = \rho_g H$ and the Alfv\`en speed is $ c_{A}^{2} = B_{0\varphi}^{2} / 4 \pi \Sigma_{g} = b_{0\varphi}^{2} / 4 \pi \rho_{g}$, with $B_{0\varphi}^{2} = H b_{0\varphi}^{2}$.

\par
\indent
The hydromagnetic equation, where no dissipation is considered, is

\begin{eqnarray}
{\partial \vec{B}_{0} \over \partial t} & = & \nabla \times ( \vec{v}_{0} 
\times \vec{B}_{0} )  \nonumber \\
& = &  ( \vec{B}_{0} \cdot \nabla ) 
\vec{v}_{0} - ( \vec{v}_{0} \cdot \nabla ) \vec{B}_{0} - \vec{B}_{0} 
( \nabla \cdot \vec{v}_{0} ) \;\; .
\label{bunpert} 
\end{eqnarray}

\noindent
The r.h.s. term of this equation is the convection term, carrying the magnetic lines of force bodily with the fluid. This equation is valid because the MF satisfy the divergence equation

\begin{equation}
\nabla \cdot \vec{B}_{0} = 0  \;\; .
\label{divb1}
\end{equation}

\noindent
In addition to the above equations, the fluid motion is completely described by adding the continuity equation

\begin{equation}
{\partial \Sigma_{g} \over \partial t} +  \nabla \cdot ( \Sigma_{g} 
\vec{v}_{0} ) = 0   \;\; ,
\label{cont1}
\end{equation}

\noindent
and the Poisson equation

\begin{equation}
\nabla^{2} \Phi_{0} = 4 \pi G \rho_{tot}  \;\; .
\label{Pois1}
\end{equation}

In the present case of a gaseous disc evolving within a halo DM, the total volume density is $\rho_{tot} = \rho_g + \rho_h $, where $ \rho_{h} $ is the volume density of the halo DM at the equatorial plane ($z=0$). The corresponding total surface density is 
$ \Sigma_{tot} = \Sigma_g + \Sigma_{dmg}$, with $\Sigma_g = \rho_g \times 2 z_g$  and $\Sigma_{dmg} = \rho_h \times 2 z_g$ for a scaleheight $z_g$. 

\par
Since we assume that the gas disc is a thin circular slab, we shall develop all the equations in cylindrical coordinates. Because we are considering only motion in the plane and the disc is axisymmetric, its properties will only vary in the radial direction $R$. We shall also assume that the components of the velocity $\vec{v}_{0}$ are $(0, v_{0\varphi}(R), 0)$ (BT08), and those of the MF are $(B_{0R}(R), B_{0\varphi}(R), 0)$. 

\section*{Appendix B: Linearized equations}
\label{sec Gperturb1}

We submit the gas disc to an external perturbation and assume that each unperturbed quantity $\mathcal{U}_{0}$ will change to a quantity $\mathcal{U} = \mathcal{U}_{0} + q$, where $q$ is an infinitesimal perturbation. In this preliminary study we look at how the baryonic component reacts to the perturbation while DM fluctuations react and contribute only to variations in the self-gravity of the disc. Replacing in (\ref{vunpert}) and considering only first order perturbations, we derive the radial and azimuthal components 

\begin{eqnarray}
{\partial v_{R} \over \partial t} & + & { v_{0\varphi} \over R } 
 {\partial v_{R} \over \partial \varphi} 
- 2 { v_{0\varphi} \over R } v_{\varphi} =  \nonumber    \\ 
& - & {1 \over \Sigma_{g}}{\partial P \over \partial R} - 
{\partial \Phi \over \partial R}
- {1 \over 4 \pi \Sigma_{g}}{\partial \over \partial R}
\left( B_{0\varphi} B_{\varphi} \right)   \nonumber   \\ 
& + & {1 \over 4 \pi \Sigma_{g}} { B_{0\varphi} \over R } 
\left\{ {\partial B_{R} \over \partial \varphi} - 2 B_{\varphi} \right\} \; ,     
\label{vlinR}
\end{eqnarray}

\begin{eqnarray}
{\partial v_{\varphi} \over \partial t} & + & v_{R} {\partial v_{0\varphi} \over 
\partial R} +  
{1 \over R}{\partial \over \partial \varphi} 
\left( v_{0\varphi} v_{\varphi} \right) 
+ { v_{0\varphi} \over R } v_{R} =    \nonumber \\   
& - & {1 \over \Sigma_{g}}{1 \over R}{\partial P \over \partial \varphi} 
- {1 \over R}{\partial \Phi \over \partial \varphi}  
+ {1 \over 4 \pi \Sigma_{g}} {B_{R} \over R}  
 {\partial \over \partial R} \left( R B_{0\varphi} \right) \; .
\label{vlinfi}
\end{eqnarray}

\noindent
The radial and azimuthal components of the linearized field equation are

\begin{equation}
\frac{\partial B_{R}}{\partial t} = \frac{ B_{0\varphi} }{ R } 
\frac{ \partial v_{R} }{ \partial \varphi } -
\frac{ v_{0\varphi} }{R } \frac{ \partial B_{R}}{ \partial \varphi } 
\label{blinR}
\end{equation}

\begin{equation}
\frac{\partial B_{\varphi}}{\partial t} =  
B_{R} \left\{ \frac{\partial v_{0\varphi}}{\partial R} - 
\frac{ v_{0\varphi} }{ R } \right\} - 
\frac{\partial}{\partial R} ( v_R B_{0\varphi} ) 
- \frac{v_{0\varphi}}{R} \frac{\partial B_{\varphi}}{\partial \varphi}  
\label{blinfi}
\end{equation}

\noindent
The divergence equation reduces to

\begin{equation}
 {1 \over R} {\partial \over \partial R} \left( R B_{R} \right) +
 {1 \over R}{\partial B_{\varphi} \over \partial \varphi}  = 0
\label{lindivb2}
\end{equation}

\noindent
and finally the continuity equation

\begin{equation}
 {\partial \Sigma_{gp} \over \partial t} + 
 {1 \over R} {\partial \over \partial R}  \left( R \Sigma_{g} v_{R} \right)
 + {1 \over R}{\partial \over \partial \varphi} \left( \Sigma_{g} v_{\varphi} 
+ \Sigma_{gp} v_{0\varphi} \right) = 0 \;\; ,
 \label{lincont2}
\end{equation}

\noindent
where $ \Sigma_{gp} $ is the perturbation of the gas surface density. To develop the dispersion equation, we take the solutions of the perturbations of the form

\begin{equation}
q(R, \varphi, t) = q_{1}(R) \exp i(m \varphi - \omega t)  \;\; .
\label{sol}
\end{equation}

\noindent
As we assume axisymmetry in the present study, $m = 0$. Because there is no simple general criterion, we shall use the WKB approximation which assumes that the long range coupling is negligible and therefore the response is determined locally. In this approximation the propagating wave has the form $\exp( ikR)$ along the $R$ direction, and we also assume that $k R \gg 1$. 
Here $\omega$ is the frequency of the oscillation and $k = 2 \pi / \lambda$ is the wave number along the radial direction (i.e. without loss of generality throughout this study we shall consider that $k > 0$). 

\par
Since the solution of the perturbed potential can be written

\begin{equation}
\Phi = \Phi_{1} \exp \{ i(k R - \omega t) - \mid kz \mid \} \;\; , 
\end{equation}

\noindent
by integrating the Poisson equation along the z-axis, one can show that the perturbed surface density is related to the perturbed gravitational potential through the relation (Binney \& Tremaine 2008 - BT08) 

\begin{equation} 
\Phi_{1} = - \frac{2 \pi G}{k} \Sigma_{tot1}  \;\; ,
\label{pot}
\end{equation}

\noindent
with $\Sigma_{tot1} = \Sigma_{g1} + \Sigma_{dm1}  = \Gamma_g \Sigma_{g1}$, and
$$ \Gamma_g = 1 + \frac{ \Sigma_{dm1} }{ \Sigma_{g1} } = 1 + \frac{ \Sigma_{dmg} }{ \Sigma_{g} } \;\; . $$

\section*{Acknowledgments}
This work is supported by a John Templeton Foundation grant.


\end{document}